# An Expansion of Polarization Control Using Semiconductor-Liquid Junctions

Peter Agbo[1,2,3]*

[1]Chemical Sciences Division
Lawrence Berkeley National Laboratory
Berkeley, CA 94720, United States

[2]Liquid Sunlight Alliance
Lawrence Berkeley National Laboratory
Berkeley, CA 94720, United States

[3]Molecular Biophysics & Integrated Bioimaging Division
Lawrence Berkeley National Laboratory
Berkeley, CA 94720, United States

*Corresponding author

**Abstract**

This report builds upon work introducing the concept of independent control over current and potential in electrocatalytic systems, as a means of improving control over their product selectivity. Previous work, describing an approach investigating independent control over potentiostat bias and current flow, implemented a biased PV-electrolyzer-type configuration. While permitting separate modulation of current and potentiostat bias, that approach precludes independent control over current flow and the applied cell potential. The present study seeks to resolve that limitation, by exploiting the Schottky diode behavior inherent to semiconductor-electrolyte interfaces. Light is explored as a prospective second degree of freedom for controlling polarization in a suitably-designed photoelectrochemical device, enabling the arbitrary control of current with respect to an applied cell potential. In stark contrast to metal electrodes, the unique property of light-dependent carrier concentrations in semiconductor electrodes forms the operative means of controlling charge fluxes at some arbitrary cell potential in PEC devices featuring a genuine semiconductor-liquid junction. This functionality carries prospects for exploring polarization states distinct from those of accessible with a dark cell, with implications for improved control over electrochemical reactions. Such opportunities are suggested by the experimental findings reported here.

The properties of traditional, dark electrochemical systems display polarization responses dictated by formalisms such as the Butler-Volmer (B-V) and Marcus-Hush-Chidsey models of electrode kinetics, with current flow scaling as the exponent of an applied cell or electrode potential (driving force)[1,2]. A key trait of these electrochemical descriptions involve majority carrier movement between a substrate and metal surface, where carrier availability, as a result of the high state occupancy of metal conduction bands, is not a factor influencing charge flow[3]. However, for serial integration of light-dependent carrier generation at a semiconductor-liquid interface, electrochemical current amplitudes will be a function of both applied potential and carrier concentration – the latter itself being a function of light intensity[4-8]. Previous work, outlining the conceptual framework for exerting increased control over current and applied potential, attempted demonstration by employing a PV-electrolyzer (PV-E) type configuration, with a photovoltaic (PV) array reverse-biased by a potentiostat being used to drive reactions in an electrochemical cell[9]. However, such a configuration precluded stabilizing the cell potential measured across the anode and cathode ohmic contacts as light intensity was changed, marking a trivial case of the light-dependent polarization behavior being sought.

Decades of effort in the fields of molecular electronics[2,10-13] and biological electron transfer theory[14-22] have demonstrated the possibility of independently inspecting the effects of driving forces and current in redox-active systems, through the imposition of tunnel junctions of varying barrier widths. The driving force and tunneling distance dependencies of charge transfer rates ($k_{ct}$) in these reactions have been shown to display characteristics captured by the semi-classical Marcus equation[14]:

$$k_{ct}(\Delta G^0, r_{DA}) = \frac{2\pi}{\hbar\sqrt{4\pi\lambda k_B T}}|H_{AB}^0 e^{(-0.5\beta r_{DA})}|^2 e^{\frac{-(\Delta G^0+\lambda)^2}{4\lambda k_B T}} \qquad \text{eq. 1.}$$

A direct consequence of these Marcus dependencies – and a key aspect of these studies setting them apart from conventional electrochemical investigations – has been their ability to investigate charge transfer rates as an independent functions of reaction driving force ($\Delta G^0$), by manipulating the distance between spatially-separated donor-acceptor sites ($r_{DA}$). Unlike parameters such as system reorganization energy ($\lambda$) and temperature ($T$), which directly influence system energetics, distance manipulation represents a unique handle for controlling kinetics without conferring simultaneous changes to reaction energetics. These relationships give rise to contours relating current and driving force, with Marcus distance-dependencies providing an additional degree of freedom for modulating current at arbitrary values of the driving force potential for a given donor-acceptor pair (Figure 1).

Physical evidence of this behavior in biochemical systems[14,23-28], and its exploitation in molecular electronics, demonstrated the general possibility of regulating current and driving forces independent of each other – a characteristic that cannot be probed through conventional polarization methods, where changes in driving force (applied potential) provide the single mode of controlling electrochemical charge transfer rates (Figure 2a). However, given that chemical reactions may invariably be decomposed into two fundamental properties – reaction kinetics (currents) and energetics (driving forces) – the ability to regulate each with an increased measure of independence could improve our ability to steer electrochemical reactivity.

This study finds its motivations through ongoing efforts to understand and control catalytic promiscuity in electrochemical $CO_2$ and $N_2$ reduction. Poor catalytic selectivity generally implies a downstream separations penalty – and therefore additional energy penalty – associated with any generated fuel. As a result, strategies to preempt such catalytic promiscuity, whether through the development of advanced catalysts or refinement of electrochemical approaches for controlling them, becomes critical. This study takes the latter approach, exploring the generalized case of light-dependent polarization, where current flow is controlled independent of the applied cell potential ($V_{app}$). The exploitation of light as an additional degree of freedom is shown to enable independent regulation of electrocatalyst kinetics and energetics in a photoelectrolyzer, impacting control over device reactivity in $CO_2$ reduction. Such control in a PEC system represents a de-facto decoupling of current density (J) and applied potential (V) in a purely functional sense, in a manner inaccessible to dark electrochemistry. Critically, the term decoupling has been adopted for brevity and is not a suggestion that fundamental J-V relationships governing circuit elements such as diodes or electrochemical cells may be in any way violated.

    Modulation of light intensity incident on the photoanode is shown to drive changes in current magnitude while keeping the applied cell potential unchanged, when applying a voltage between the cathode and the front contact of the photoanode. The result is a construct where measured current and applied cell potential can be varied independent of one another, with the measured polarization response marking a convolution of both PV and electrocatalyst characteristics. Photocurrent, and therefore polarization curves of the photoelectrochemical (PEC) electrolyzer, vary as functions of the light intensity incident on the photoanode (Figure 2b-c). Continuous variation of light incident on the photoelectrolyzer generates a continuum of these light-dependent, J-V response curves (Figure 2d), with polarization marking a composition of both electrode kinetics and the light-dependent diode behavior of the semiconductor-liquid junction. Light sensitivity was incorporated through the use of n-type, Si <100> photoelectrodes, modified with Ni catalyst layers,[29,30] as light-sensitive anodes for gating current flow in a compact photoelectrolyzer (Figure 3a).

    It was previously proposed that regions where current and potential may be arbitrarily controlled in a light-coupled system are captured by the expression[9]:

$$\int_{V_0}^{V_1} J(\phi_{max}, V_{app}) dV \qquad \text{(expression 1)},$$

where $J(\phi_{max}, V_{app})$ denotes the cell current density under maximum illumination. The same expression is shown to be descriptive of current flow under the control of photoelectrodes whose catalytic layers represent genuine, Schottky-type semiconductor-liquid junctions, rather than buried junctions governed solely by the bulk (> 20 nm) properties of an overlaying metal catalyst. In the context of cathodes, which are often designed with various reductive chemistries in mind, such a device provides a strategy for testing the effects of arbitrarily setting current with respect to applied potential. This opens a path for controlling carrier kinetics independently of carrier energetics – a measure of control disallowed by the polarization constraints of dark voltammetry. While implementation of this concept using a photoelectrochemical cathode as the current-limiting element – in either two or three electrode cell configurations – should be possible, doing either requires a candidate catalyst that can be reliably constituted on a semiconductor at thicknesses small enough to prevent conversion of the semiconductor-liquid junction to buried junction behavior. This requirement is not to be universally applicable for all catalysts and, where possible, remains highly dependent on fabrication procedures. However, using light-sensitive photoanodes to effectively gate current at dark cathodes in a two-electrode configuration, circumvents these complications, and allows exploration of decoupling phenomena using any catalyst that can be constituted on dark cathodes.

    A cell for demonstrating light-dependent, electrochemical decoupling was assembled through the fabrication of a photoactive anode doubling as an OER catalyst and point of light absorption (Figures 3b-d). Si<100> (0.1-0.5 ohm-cm) chips were modified with 20 nm of Ti serving as ohmic contacts to the unpolished side of the Si substrates (Ti back contact). Afterwards, the polished side of the Si wafers (retaining their native oxide) were amended with a 2 nm Ni catalyst layer, deposited via radio frequency metal sputtering (S.1). Afterwards, a busbar composed of 100 nm Au was sputtered on the Ni side of the n-Si/SiO$_2$ substrates (Au front contact)[30,31]. Ta leads were contacted to the Ti layer and Au busbar to serve as back and front contacts to the modified Si anodes, and were each protected from solution exposure with epoxy. These electrodes were mounted in a compact, transparent electrochemical cell containing an anion exchange membrane separator and 100 nm Au catalyst/Toray carbon paper cathodes (S.1,

S.2). Light illumination was focused on through the anode compartment, with a variable filter between anode and source light controlling incident photon flux.

Biasing this cell between the photoanode's Ti back contact and cathode (configuration ii) displays the light dependent polarization typical of a photoelectrolyzer, with a negative shift (60-200 mV) in bias response observed as a result of the photovoltage generated across the semiconducting anode[5-7]. However, independent voltage sensing of the cell potential between the Au front anode contact and the device cathode, is found to be unstable as a function of changing the light intensity (and current)[9,32]. By contrast, voltage application between the photoanode's Au surface contact and the dark cathode results in a current response with significantly reduced shifts in junction potential as a function of light intensity. In this configuration (i), measurement of the applied cell potential is found to remain stable as current is arbitrarily switched by changing the intensity of photoanodic insolation.

While Au should be expected to form a Schottky diode on n-Si/SiO$_2$ substrates, measurement of the I-V properties of these photoanode chips by polarizing between the front and back contact reveal a device behavior diagnostic of a mixed tunneling/Schottky contact (Figure 3b), consistent with findings by Wang et al[33]. Photoanode I-V curve measurements allow for probing potential drops between the Au contact and the n-Si substrate. The non-rectifying branch of the I-V curve represents the directionality of majority carrier flow during actual device operation and yields a resistance of 30 Ω. This suggests that the Au/SiO$_2$/Si tunnel junction is a major contributor to ohmic losses in this device. Critically, a light-sensitive response is only present in the I-V curve for the reverse-bias case of the Au/SiO$_2$/n-Si junction, indicating that photon-induced shifts in n-Si photoconductivity are not responsible for the light-dependent response of these photoanodes under electrocatalytic conditions. Cell testing with biasing the Au contact in both wet and insulated modalities are found to preserve light-sensitive polarization derived from the n-Si/SiO$_2$/Ni/NiO$_X$-electrolyte junction.

Thickness of the Ni catalyst on the Si photoanode is found to be a critical determinant of device behavior. At 2 nm, Ni thickness is such that the photoanode-liquid junction constitutes a genuine photoelectrochemical interface with Schottky diode characteristics, rather than forming a buried junction beneath a contiguous catalyst layer displaying bulk metallic character. This distinction is key, as over-depositing Ni on the n-Si/SiO$_2$ substrate will fail to establish the anode catalyst-liquid junction as a genuine, in-situ diode contact. Instead, a thick metal layer with bulk metal electrical properties will result in device behavior that is indistinguishable from a biased PV-electrolyzer, a configuration where changes in illumination profile will simply yield polarization curves corresponding to the dark cell (or working electrode in 3-electrode cases). Previous reports by Kenney et al[29]. noted that for catalyst thicknesses of 5 nm and greater, behavior of the n-Si/SiO$_2$/NiO$_X$/Ni assumes character of an MIS solar cell with a metallic Ni layer serving as the catalyst. At thicknesses significantly greater (> 10 nm), the Ni catalyst layer is essentially contiguous, thick enough to form an effective buried junction[29]. At these greater thicknesses, biasing the integrated photoelectrolyzer in configuration (i) gives a polarization response that is unresponsive to light illumination, with the Ni layer behaving as a dark metal catalyst, as evidenced from predictive simulations (S.3) and experiment using a 20 nm Ni layer (S.9).

Photovoltage generated at the semiconductor-liquid junction ($V_{Lj}$) will be set by the difference in electrochemical potential between the semiconductor Fermi level and equilibrium speciation of the dissolved redox-active couple interacting with the semiconducting anode, according to:

$$V_{Lj} = E_F - [E^0_{A/A^-} + RT \ln(A^-/A)] \qquad \text{eq. 2}^{3,34},$$

where $E_F$ is the semiconductor Fermi level and here, A-/A denotes the H$_2$O/O$_2$ redox couple. In the case of a two-electrode cell, as used in this study, driving forces for the net cell reaction are given by the overall electrochemical cell potential:

$$V = V_{app} + V_{Lj}(\phi) \qquad \text{eq. 3.}$$

The driving force for electrochemical reactions, V, is given by equation 3, where $V_{Lj}$ is a floating quantity free to shift as changes in light intensity (ϕ) alter the magnitude of current flow through the cell[35]. In configuration (ii), changes in light intensity induce undesired changes to the junction potential, altering cell reaction driving forces. Such effects are less pronounced in the case of a (front) contact located at the semiconductor-liquid junction, with applied bias compensating for fluctuations in $V_{lj}$. The light-mediated shifts in cell potential, as measured between the cathode and a photoanode-liquid interface, move in accordance with the behavior of semiconductor-liquid junctions given by rearrangement of the Shockley-Diode equation (eq. 4)[4,7]:

$$V_{Lj}(\phi) = \frac{nk_B T}{q} \ln\left(\frac{J(V,\phi) + J(\phi)_{sc}}{J_0} + 1\right) \quad \text{eq. 4.}$$

Here, the value of the photovoltage generated at the photoanode, and so the measured value of the potential measured across the cell, shifts in proportion to the logarithm of the cell current as light incident on the photoanode – and therefore cell current – is modulated[5,7,30,36]. Notably, while current can be arbitrarily tuned with respect to the applied potential (measured between the cathode and anode busbar), measuring the precise value of the photoanode surface potential with respect to the cathode is a separate problem, made difficult because of the complexity of contacting a spatially-diffuse catalyst layer[30]. However, given the distinct polarization characteristics between a Schottky liquid junction (diode behavior) vs a buried junction (governed by dark, B-V-type polarization), even real shifts in junction potential should permit the exploration of device polarization states that are markedly distinct from those accessible by a dark device, PV-EC device, or buried junction PEC device, with potential downstream consequences for influencing product selectivity.

The particular cases of PV-EC devices and PEC cells with buried junctions, yield shifts in polarization behavior defined by the B-V characteristics of the dark catalysts comprising the respective photoactive cells. Attempts at bias application at the wet junction of a buried junction device does not enable light-dependent modulation of current in any respect, as the bulk-metal characteristics of buried-junction PEC devices make their metallized interfaces insensitive to photoillumination, unless a back-contact is used (S.9)[32]. However, in the case of the PEC device presented here, bias application between the anode front contact and cathode (configuration i) does preserve light-responsiveness, the consequence of a semiconductor-liquid junction displaying genuine Schottky diode behavior.

Modification of a previously derived analytical model for PV-electrolyzer polarization[9] remains relevant to the case of photoelectrochemical decoupling using semiconductor-liquid interfaces, capable of modeling the observed J-V response of the integrated photoelectrolyzer:

$$J(\phi,V) = \frac{n_e F}{N_A} k_2 k_{ET}\left(1 - \frac{k'_{ET}}{k'_{ET}+k_2}\right) \times \frac{[A]_T + \frac{k_{PV}(\phi,V)}{k'_{ET}}f}{(k'_{ET}+k_2)\left(\frac{1}{k_1[OH^-]}\left(k'_1 + \frac{k_2 k_{ET}}{k_2+k'_{ET}}\right) + 1 + \frac{k_{ET}}{k'_{ET}}\right)} \quad \text{eq. 5a.}$$

Parameterization for terms $k_1$, $k_1'$, $k_2$, and $k_2'$ are as described in the Supporting Information. Rates $k_{ET}$ and $k_{ET}'$ represent back and forward electron transfer rate constants described by the B-V equation (S.1). Active site concentration at the current-limiting electrode is represented by $[A]_T$ ($10^{14}$ cm$^{-2}$), while $f$ is a composite term forcing the entire function to zero in the limit of zero active centers[9]. The variable $n_e$ gives the number of electrons transferred per reaction (2), $F$ is faraday's constant, and $N_A$ is avagadro's number. The term $k_{pv}$ gives the Shockley diode equation, rewritten as a unimolecular rate constant (s$^{-1}$):

$$k_{pv}(\phi,V) = \frac{N_A}{n_e F}[I(\phi)_{sc} - I_0\exp(\frac{q(V-IR_s)}{n_d k_B T}) - \frac{V-IR_s}{R_{sh}}] \qquad \text{eq. 5b,}$$

where the ideality factor, $n_d$, is taken as 1, and an $I_0$ of 0.1 nA was used for simulation. The single-diode equation used here accounts for current attenuation resulting from series resistances ($R_s$) and shunt resistances, $R_{sh}$[37]. Practical issues arise from the form of equation 5b, which requires knowledge of the very current being calculated, as an input parameter. As a result, Equation 5b was calculated numerically using an iterative, Newton's method-type approach (S.3). Using equations 5a and 5b, it was possible to reproduce device physics unique to a front-contacted, photoelectrode device (Figures 4a-d). In particular, calculated polarization traces captures key changes in the observed J-V curve shape, which transitions from a slow exponential rise weakly bounded by $J_{sc}$ at high illumination powers, to a curve that strictly plateaus at values of $J_{sc}$ at low illumination powers. Modeling suggests these observations are consistent with a photoelectrochemical device governed primarily by the kinetics of a front-contacted photoanode featuring a high shunt resistance ($R_{sh}$ > 15 kΩ) and moderate series resistance ($R_s$ ~ 57 Ω). In addition to the ohmic resistance associated with charge tunneling, the requirement that the photoanode be illuminated through the solution-exposed side results in a catalyst layer positioned opposite, rather than against, the cell membrane. The results are high ionic transport path lengths that significantly raise the overall cell resistance, $R_\Omega$ (38-57 Ω, S.10) relative to a typical membrane-electrode-assembly type configuration.

Measurements of light-dependent device polarization evidence a clear sensitivity to photon flux incident on the photoanode, both in cases of cathode polarization with respect to either the Au front (solution) or Ti back photoanode contacts. Cathode polarization versus the photoanode backcontact (configuration ii) while using the Au contact for voltage sensing only, demonstrates that light variation using this mode of connectivity results in significant voltage drops between the anode catalyst and cathode that vary with light intensity (Figure 5a). However, stabilizing the potential measured between the front contact and the cathode is made possible by bias application between the cathode and photoanode's Au front contact at the semiconductor-liquid interface (configuration i, Figures 5b,c).

The details of photoanode fabrication prove crucial to demonstrating the behavior and physical principle of current-potential decoupling embodied by configuration (i). In particular, the conductivity of the catalyst layer deposited on the semiconductor is found to be a key determinant of whether functional decoupling of current and applied potential can be realized in practice. Charge collection using a front contact at the solution-catalyst interface results in the possibility of two paths for charge flow between catalyst sites and the Au contact to be established: at sufficiently thin catalyst layers, conductivity of the Ni layer is too low (since catalyst sites are too physically isolated) to enable free charge movement to the Au contact without proceeding through the underlying semiconductor. However, significantly thicker Ni layers (> 20 nm) are found to be conductive enough to form an electrical shunt, allowing charge to bypass movement through the semiconductor and instead, move along electrically-contiguous paths of Ni deposits between the Au contact and initial points of water oxidation. This is readily observed by inspecting polarization behavior in the limit of thick Ni layers when operating in configuration (i) (S.9). Modeling behavior with a high shunt resistance (thin catalyst layers) results in polarization curves that feature low dark currents, as a result of insufficient carrier availability in the semiconductor, and the absence of parallel conduction through the catalyst layer only (Figure 4b, S.3). However, reducing the photoanode's shunt resistance (by establishing thicker catalyst layers) increases the dark current, by introducing a parallel path for charge flow that bypasses the semiconductor. Notably, this difference in polarization behavior does not arise for the cases of thin and thick catalyst layers when the photoanode is contacted through the Ti back contact. In this case, the relative orientation between the contact and the semiconductor and catalyst layers form a serial configuration, with charge extracted at catalyst sites being required to flow through the semiconductor (and is therefore limited by light-dependent, minority carrier concentrations in the semiconductor), irrespective of the catalyst layer thickness. As a result, polarization in configuration (ii) displays low dark currents for all catalyst layer thicknesses, consistent with reports throughout PEC literature[6,7,29–31,36,38–45], where light-sensitivity is observed for back-contacted silicon photoelectrodes featuring several tens of nanometers of catalyst.

Testing product evolution of the photoelectrolyzer in configuration (i), when using 100 nm Au on Toray carbon paper as a cathode under 5 sccm $CO_2$ flow, results in significant changes to the $CO/H_2$ product distribution as light intensity is changed at a single value of the applied cell potential (Figure 6a, S.4-5, S.7). Specifically, changes to light-limited currents spanning 1-6 mA at 4.0 V cell potential, are shown to cause concurrent changes to the $CO/H_2$ ratio, with CO fractions increasing at lower current densities, and jumping significantly at 1 mA. Similar behavior is observed at 3.5 and 3.0 V (S.5). Notably, data acquisition over multiple current levels at a single applied cell potential demonstrate that product distributions are not unique functions of the applied potential, for a given photoelectrochemical cell. Instead, the descriptor displaying the clearest relationship to product distribution appears to be device power, a metric that effectively collapses the information represented by the kinetics (J) and energetics (V) into a single unit, JxV (Figures 6b,c, S.6). In the particular case of the devices tested, observed trends in power density seem to be driven by the current dependence of the product distributions, which show similar monotonic dependencies mirroring those of the power function (S.6).

Light-dependent electrochemical impedance spectroscopy conducted of the system in configuration (i) shows that the value of the ohmic resistance ($R_\Omega$) as measured in the high-frequency regime, remains unchanged as a function of light intensity. This is distinct from impedance behavior observed by varying the value of the ohmic resistance through placement of resistive elements in series with the Au contact, causing the magnitude of $R_\Omega$ to increase with increasing values of the external resistance (S.11). These results are key, demonstrating that current modulation as a function of varied illumination is not the result of changes in $iR$ losses across the cell due to changes in photoconductivity of the n-Si anode. Instead, changes in light intensity coincide with shifts in a secondary arc, assigned to charge transfer impedance (S.10). The finding agrees well with the basic premise that the light-dependent availability of carriers in semiconductors – in contrast with the excess of available majority carriers in metal conduction bands – should provide a way of limiting carrier flow in a manner distinct from dissipative $iR$ drop. However, at sufficiently large values of $R_\Omega$ and/or device current, $iR$ drops can force undesired changes in real cell potential, representing important factors that must be considered when employing this approach. In this particular study, $iR$ compensation of these data are still found to yield the key power and potential dependencies observed for $CO/H_2$ distributions (S.8). Future improvements in device design will focus on mitigating ohmic losses through improved design of the Au contact, reducing ionic path lengths – potentially through the use of back-illuminated photoanodes[46,47] – and using KOH rather than the lower conductivity borate-based electrolyte used in this study[29].

These results, in conjunction with the ability to alter product distributions at a given applied cell potential through arbitrary changes in cell current density, demonstrate the basic viability – and potential utility – of using light to expand control over electrochemical polarization. Notably, while the system designed here effectively allows arbitrary movement of current with respect to applied potential (within the constraints set by expression 1), this provides little information on the electrochemical potential profile as measured between the photoanode catalyst sites and cathode. However, the unique polarization behavior of semiconductor-liquid diodes display surface potentials that, unlike pure metal-liquid junctions, show dependencies on applied potential allowing electrode response to be influenced by factors, such as Fermi level pinning by surface states[48]. These relationships result in surface potentials whose polarization dependencies are likely to be markedly distinct from those of the corresponding bulk metal catalyst. As a result, PEC device implementations such as the one detailed here, should at minimum, enable us to access polarization states distinct from those accessible with mere bulk metal catalysts, with the possibility of steering product distributions by manipulating parameters other than applied voltage.

**Materials and Methods**

Materials and methods for this study can be found in the Supplementary Information

**References**

(1) Butler, J. a. V. The Mechanism of Overvoltage and Its Relation to the Combination of Hydrogen Atoms at Metal Electrodes. *Trans. Faraday Soc.* **1932**, *28* (0), 379–382. https://doi.org/10.1039/TF9322800379.

(2) Chidsey, C. E. D. Free Energy and Temperature Dependence of Electron Transfer at the Metal-Electrolyte Interface. *Science* **1991**, *251* (4996), 919–922. https://doi.org/10.1126/science.251.4996.919.

(3) Bockris, J. O.; Reddy, A. K. N.; Gamboa-Aldeco, M. *Modern Electrochemistry*; Springer, 2000.

(4) Shockley, W.; Queisser, H. J. Detailed Balance Limit of Efficiency of P-n Junction Solar Cells. *Journal of Applied Physics* **1961**, *32* (3), 510–519. https://doi.org/10.1063/1.1736034.

(5) Gerischer, H. On the Stability of Semiconductor Electrodes against Photodecomposition. *Journal of Electroanalytical Chemistry and Interfacial Electrochemistry* **1977**, *82* (1), 133–143. https://doi.org/10.1016/S0022-0728(77)80253-2.

(6) Jiang, C.; A. Moniz, S. J.; Wang, A.; Zhang, T.; Tang, J. Photoelectrochemical Devices for Solar Water Splitting – Materials and Challenges. *Chemical Society Reviews* **2017**, *46* (15), 4645–4660. https://doi.org/10.1039/C6CS00306K.

(7) Walter, M. G.; Warren, E. L.; McKone, J. R.; Boettcher, S. W.; Mi, Q.; Santori, E. A.; Lewis, N. S. Solar Water Splitting Cells. *Chem. Rev.* **2010**, *110* (11), 6446–6473. https://doi.org/10.1021/cr1002326.

(8) Fujishima, A.; Honda, K. Electrochemical Photolysis of Water at a Semiconductor Electrode. *Nature* **1972**, *238* (5358), 37–38. https://doi.org/10.1038/238037a0.

(9) Agbo, P. J–V Decoupling: Independent Control over Current and Potential in Electrocatalysis. *J. Phys. Chem. C* **2020**, *124* (52), 28387–28394. https://doi.org/10.1021/acs.jpcc.0c08142.

(10) Smalley, J. F.; Feldberg, S. W.; Chidsey, C. E. D.; Linford, M. R.; Newton, M. D.; Liu, Y.-P. The Kinetics of Electron Transfer Through Ferrocene-Terminated Alkanethiol Monolayers on Gold. *J. Phys. Chem.* **1995**, *99* (35), 13141–13149. https://doi.org/10.1021/j100035a016.

(11) Neuhausen, A. B.; Hosseini, A.; Sulpizio, J. A.; Chidsey, C. E. D.; Goldhaber-Gordon, D. Molecular Junctions of Self-Assembled Monolayers with Conducting Polymer Contacts. *ACS Nano* **2012**, *6* (11), 9920–9931. https://doi.org/10.1021/nn3035183.

(12) Devaraj, N. K.; Decreau, R. A.; Ebina, W.; Collman, J. P.; Chidsey, C. E. D. Rate of Interfacial Electron Transfer through the 1,2,3-Triazole Linkage. *J. Phys. Chem. B* **2006**, *110* (32), 15955–15962. https://doi.org/10.1021/jp057416p.

(13) Finklea, H. O.; Hanshew, D. D. Electron-Transfer Kinetics in Organized Thiol Monolayers with Attached Pentaammine(Pyridine)Ruthenium Redox Centers. *J. Am. Chem. Soc.* **1992**, *114* (9), 3173–3181. https://doi.org/10.1021/ja00035a001.

(14) Marcus, R. A.; Sutin, N. Electron Transfers in Chemistry and Biology. *Biochimica et Biophysica Acta (BBA) - Reviews on Bioenergetics* **1985**, *811* (3), 265–322. https://doi.org/10.1016/0304-4173(85)90014-X.

(15) Gray, H. B.; Winkler, J. R. Electron Flow through Proteins. *Chemical Physics Letters* **2009**, *483* (1–3), 1–9. https://doi.org/10.1016/j.cplett.2009.10.051.

(16) Winkler, J. R.; Di Bilio, A.; Farrow, N. A.; Richards, J. H.; Gray, H. B. Electron Tunneling in Biological Molecules. *Pure and Applied Chemistry* **1999**, *71* (9), 1753–1764. https://doi.org/10.1351/pac199971091753.

(17) Gray, H. B.; Winkler, J. R. Long-Range Electron Transfer. *PNAS* **2005**, *102* (10), 3534–3539. https://doi.org/10.1073/pnas.0408029102.

(18) Warren, J. J.; Ener, M. E.; Vlček Jr., A.; Winkler, J. R.; Gray, H. B. Electron Hopping through Proteins. *Coordination Chemistry Reviews* No. 0. https://doi.org/10.1016/j.ccr.2012.03.032.

(19) Ponce, A.; Gray, H. B.; Winkler, J. R. Electron Tunneling through Water: Oxidative Quenching of Electronically Excited Ru(Tpy)$_2$$^{2+}$ (Tpy = 2,2':6,2''-Terpyridine) by Ferric Ions in Aqueous Glasses at 77 K. *J. Am. Chem. Soc.* **2000**, *122* (34), 8187–8191. https://doi.org/10.1021/ja000017h.

(20) Wenger, O. S.; Leigh, B. S.; Villahermosa, R. M.; Gray, H. B.; Winkler, J. R. Electron Tunneling Through Organic Molecules in Frozen Glasses. *Science* **2005**, *307* (5706), 99–102. https://doi.org/10.1126/science.1103818.

(21) Wenger, O. S. Barrier Heights in Long-Range Electron Tunneling. *Inorganica Chimica Acta* **2011**, *374* (1), 3–9. https://doi.org/10.1016/j.ica.2011.01.058.

(22) Wenger, O. S. How Donor–Bridge–Acceptor Energetics Influence Electron Tunneling Dynamics and Their Distance Dependences. *Acc. Chem. Res.* **2010**, *44* (1), 25–35. https://doi.org/10.1021/ar100092v.

(23) Marcus, R. A. Electron Transfer Theory and Its Inception. *Physical Chemistry Chemical Physics* **2012**. https://doi.org/10.1039/c2cp90116a.

(24) Gray, H. B.; Winkler, J. R. Electron Tunneling Through Proteins. *Quarterly Reviews of Biophysics* **2003**, *36* (03), 341–372. https://doi.org/10.1017/S0033583503003913.

(25) Gray, H. B.; Winkler, J. R. Electron Flow through Metalloproteins. *Biochimica et Biophysica Acta (BBA) - Bioenergetics* **2010**, *1797* (9), 1563–1572. https://doi.org/10.1016/j.bbabio.2010.05.001.

(26) Beratan, D. N.; Betts, J. N.; Onuchic, J. N. Tunneling Pathway and Redox-State-Dependent Electronic Couplings at Nearly Fixed Distance in Electron Transfer Proteins. *J. Phys. Chem.* **1992**, *96* (7), 2852–2855. https://doi.org/10.1021/j100186a014.

(27) Beratan, D. N.; Betts, J. N.; Onuchic, J. N. Protein Electron Transfer Rates Set by the Bridging Secondary and Tertiary Structure. *Science* **1991**, *252* (5010), 1285–1288. https://doi.org/10.1126/science.1656523.

(28) Balabin, I. A.; Beratan, D. N.; Skourtis, S. S. Persistence of Structure Over Fluctuations in Biological Electron-Transfer Reactions. *Phys. Rev. Lett.* **2008**, *101* (15), 158102. https://doi.org/10.1103/PhysRevLett.101.158102.

(29) Kenney, M. J.; Gong, M.; Li, Y.; Wu, J. Z.; Feng, J.; Lanza, M.; Dai, H. High-Performance Silicon Photoanodes Passivated with Ultrathin Nickel Films for Water Oxidation. *Science* **2013**, *342* (6160), 836–840. https://doi.org/10.1126/science.1241327.

(30) Laskowski, F. A. L.; Nellist, M. R.; Venkatkarthick, R.; Boettcher, S. W. Junction Behavior of N-Si Photoanodes Protected by Thin Ni Elucidated from Dual Working Electrode Photoelectrochemistry. *Energy Environ. Sci.* **2017**, *10* (2), 570–579. https://doi.org/DOI: 10.1039/c8cc08146h.

(31) Lee, S. A.; Park, I. J.; Yang, J. W.; Park, J.; Lee, T. H.; Kim, C.; Moon, J.; Kim, J. Y.; Jang, H. W. Electrodeposited Heterogeneous Nickel-Based Catalysts on Silicon for Efficient Sunlight-Assisted Water Splitting. *Cell Reports Physical Science* **2020**, *1* (10), 100219. https://doi.org/10.1016/j.xcrp.2020.100219.

(32) Yap, K. M. K.; Lee, S.-W.; Steiner, M. A.; Avilés Acosta, J. E.; Kang, D.; Kim, D.; Warren, E. L.; Nielander, A. C.; Jaramillo, T. F. A Framework for Understanding Efficient Diurnal CO$_2$ Reduction Using Si and GaAs Photocathodes. *Chem Catalysis* **2023**, 100641. https://doi.org/10.1016/j.checat.2023.100641.

(33) Wang, F.; Liu, Y.; Duffin, T. J.; Kalathingal, V.; Gao, S.; Hu, W.; Guo, Y.; Chua, S.-J.; Nijhuis, C. A. Silicon-Based Quantum Mechanical Tunnel Junction for Plasmon Excitation from Low-Energy Electron Tunneling. *ACS Photonics* **2021**, *8* (7), 1951–1960. https://doi.org/10.1021/acsphotonics.0c01913.

(34) Bard, A. J.; Faulkner, L. R. *Electrochemical Methods: Fundamentals and Applications*; Wiley, 2001.

(35) Wang, Z.; Lyu, M.; Chen, P.; Wang, S.; Wang, L. Energy Loss Analysis in Photoelectrochemical Water Splitting: A Case Study of Hematite Photoanodes. *Phys. Chem. Chem. Phys.* **2018**, *20* (35), 22629–22635. https://doi.org/10.1039/C8CP04021D.





(36) Rosenbluth, M. L.; Lewis, N. S. "Ideal" Behavior of the Open Circuit Voltage of Semiconductor/Liquid Junctions. *J. Phys. Chem.* **1989**, *93* (9), 3735–3740. https://doi.org/10.1021/j100346a072.

(37) Ben Or, A.; Appelbaum, J. Estimation of Multi-Junction Solar Cell Parameters. *Progress in Photovoltaics: Research and Applications* **2013**, *21* (4), 713–723. https://doi.org/10.1002/pip.2158.

(38) Zhu, T.; Chong, M. N. Prospects of Metal–Insulator–Semiconductor (MIS) Nanojunction Structures for Enhanced Hydrogen Evolution in Photoelectrochemical Cells: A Review. *Nano Energy* **2015**, *12*, 347–373. https://doi.org/10.1016/j.nanoen.2015.01.001.

(39) Chen, Y. W.; Prange, J. D.; Dühnen, S.; Park, Y.; Gunji, M.; Chidsey, C. E. D.; McIntyre, P. C. Atomic Layer-Deposited Tunnel Oxide Stabilizes Silicon Photoanodes for Water Oxidation. *Nature Mater* **2011**, *10* (7), 539–544. https://doi.org/10.1038/nmat3047.

(40) Cordova, I. A.; Peng, Q.; Ferrall, I. L.; Rieth, A. J.; Hoertz, P. G.; Glass, J. T. Enhanced Photoelectrochemical Water Oxidation via Atomic Layer Deposition of $TiO_2$ on Fluorine-Doped Tin Oxide Nanoparticle Films. *Nanoscale* **2015**, *7* (18), 8584–8592. https://doi.org/10.1039/C4NR07377K.

(41) Noh, E.; Noh, K.-J.; Yun, K.-S.; Kim, B.-R.; Jeong, H.-J.; Oh, H.-J.; Jung, S.-C.; Kang, W.-S.; Kim, S.-J. Enhanced Water Splitting by $Fe_2O_3$-$TiO_2$-FTO Photoanode with Modified Energy Band Structure. *The Scientific World Journal* **2013**, *2013*, e723201. https://doi.org/10.1155/2013/723201.

(42) Zhou, W.; Fan, R.; Hu, T.; Huang, G.; Dong, W.; Wu, X.; Shen, M. 5.1% Efficiency of Si Photoanodes for Photoelectrochemical Water Splitting Catalyzed by Porous NiFe (Oxy)Hydroxide Converted from NiFe Oxysulfide. *Chemical Communications* **2019**, *55* (84), 12627–12630. https://doi.org/10.1039/C9CC06413C.

(43) Huang, G.; Fan, R.; Zhou, X.; Xu, Z.; Zhou, W.; Dong, W.; Shen, M. A Porous Ni-O/Ni/Si Photoanode for Stable and Efficient Photoelectrochemical Water Splitting. *Chem. Commun.* **2019**, *55* (3), 377–380. https://doi.org/10.1039/C8CC08146H.

(44) Nakajima, T.; Hagino, A.; Nakamura, T.; Tsuchiya, T.; Sayama, K. $WO_3$ Nanosponge Photoanodes with High Applied Bias Photon-to-Current Efficiency for Solar Hydrogen and Peroxydisulfate Production. *J. Mater. Chem. A* **2016**, *4* (45), 17809–17818. https://doi.org/10.1039/C6TA07997K.

(45) Liu, R.; Yuan, G.; Joe, C. L.; Lightburn, T. E.; Tan, K. L.; Wang, D. Silicon Nanowires as Photoelectrodes for Carbon Dioxide Fixation. *Angewandte Chemie International Edition* **2012**, n/a-n/a. https://doi.org/10.1002/anie.201202569.

(46) Liu, B.; Wang, T.; Wang, S.; Zhang, G.; Zhong, D.; Yuan, T.; Dong, H.; Wu, B.; Gong, J. Back-Illuminated Photoelectrochemical Flow Cell for Efficient $CO_2$ Reduction. *Nat Commun* **2022**, *13* (1), 7111. https://doi.org/10.1038/s41467-022-34926-x.

(47) Bae, D.; Pedersen, T.; Seger, B.; Malizia, M.; Kuznetsov, A.; Hansen, O.; Chorkendorff, I.; Vesborg, P. C. K. Back-Illuminated Si Photocathode: A Combined Experimental and Theoretical Study for Photocatalytic Hydrogen Evolution. *Energy Environ. Sci.* **2015**, *8* (2), 650–660. https://doi.org/10.1039/C4EE03723E.

(48) Lichterman, M. F.; Hu, S.; Richter, M. H.; Crumlin, E. J.; Axnanda, S.; Favaro, M.; Drisdell, W.; Hussain, Z.; Mayer, T.; Brunschwig, B. S.; Lewis, N. S.; Liu, Z.; Lewerenz, H.-J. Direct Observation of the Energetics at a Semiconductor/Liquid Junction by Operando X-Ray Photoelectron Spectroscopy. *Energy Environ. Sci.* **2015**, *8* (8), 2409–2416. https://doi.org/10.1039/C5EE01014D.


**Materials and Correspondence**

Correspondence related to this study should be sent to pagbo@lbl.gov.

**Supplementary Information**

Supplementary Information are available for this manuscript.

**Competing Interests**




The author declares no competing interests.

**ORCID**

http://orcid.org/0000-0003-3066-4791

**Acknowledgements**

The author would like to acknowledge Drs. Joseph Varghese, Tobias Kistler, Walter Drisdell, Gurudayal Singhal and Joel Ager for their helpful comments. This work was funded by the Liquid Sunlight Alliance, a DOE Innovation Hub, under award number DE-SC0004993.


**Figures**

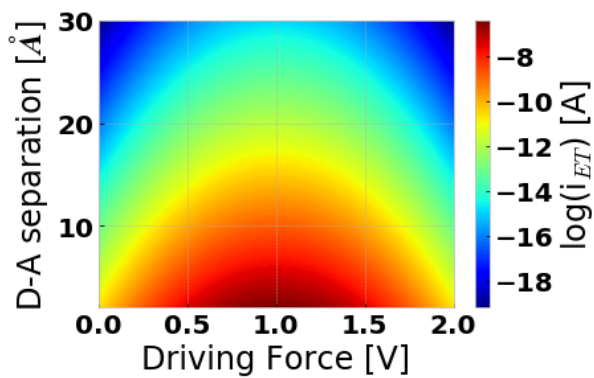

**Figure 1**

Marcus relations of tunneling currents and driving force display multiple degrees of freedom, including donor-acceptor (D-A) distance-dependence, according to eq. 1. As a result, tunneling current may be modulated over several orders of magnitude at a single value of the driving force, by controlling D-A separation. Contours shown are calculated using the semi-classical Marcus equation for the following canonical Marcus parameters: $k_0 = 10^{13}$ s$^{-1}$, $\lambda = 1.0$ eV, $\beta = 14$ nm$^{-1}$.



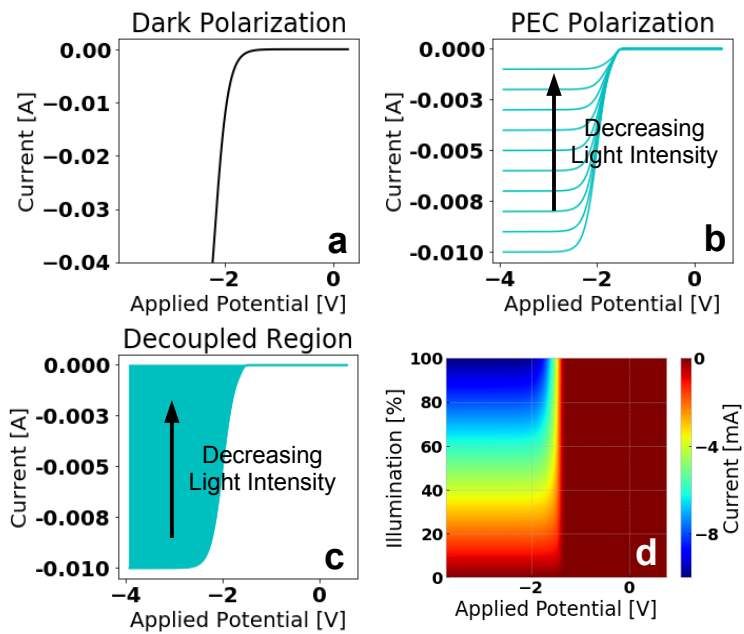

**Figure 2**
(a) Simulated dark cell polarization governed solely by B-V kinetics. (b, c) An idealized simulation of PEC electrolyzer polarization under light using equations 5a and 5b. Arbitrary changes to light intensity enable a wider range of J-V curves to be accessed (V is the applied cell voltage). (d) Simulated contours of independent modulation of applied cell potential and current. Here, introduction of light serves as a secondary handle for controlling current, acting as a surrogate for the distance-dependence observed in Marcus-type tunneling currents.



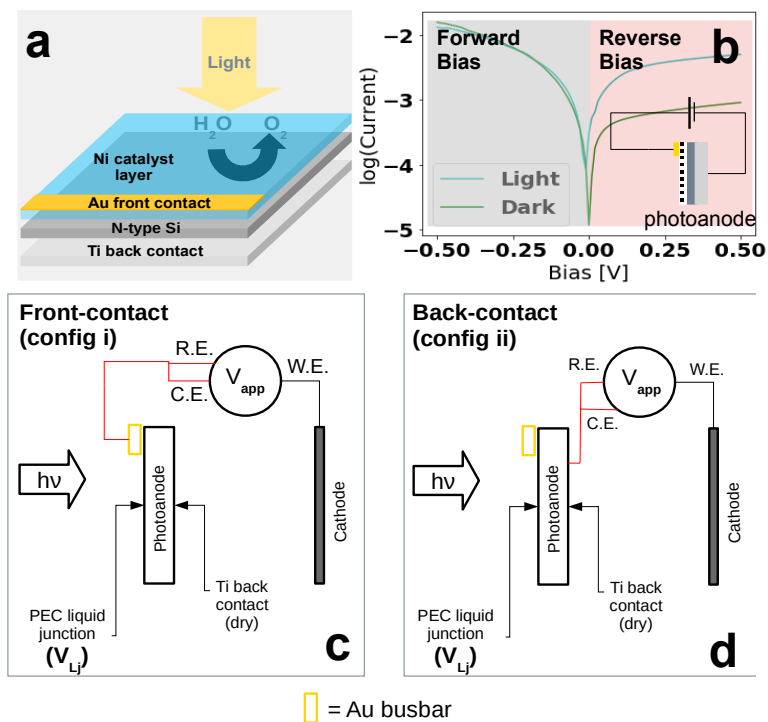

□ = Au busbar

**Figure 3**
**(a)** Construction of the Si photoelectrode used as the cell photoanode. Biasing versus a dark cathode may be made through either the photoanode's traditional back contact (Ti) or a front contact at the semiconductor-liquid junction (Au busbar). **(b)** Dry I-V curves of the photoanode chips under light and dark. Under forward bias (left branches), the configuration relevant to electrochemical cell operation, no light dependence is observed and current response is high, suggestive of a junction of mixed tunneling and Schottky barrier character. **(c)** The 2-electrode cell configuration (i) is found to enable functional decoupling through application of an applied cell voltage between the front (solution) contact of a photoanode at an Au busbar, and a cathode. **(d)** In the 2-electrode cell configuration (ii), bias is applied between the photoanode back contact (dry Ti layer) and the cathode. As a result, the photovoltage generated at the anode is a floating quantity that moves as a function of the light intensity and spectrum incident on the photoanode; attempting to adjust current flow through the cell at a specific bias, by modulating anode illumination, will cause the real cell potential – a partial function of the photoanode voltage – to shift in accordance with equation (3).



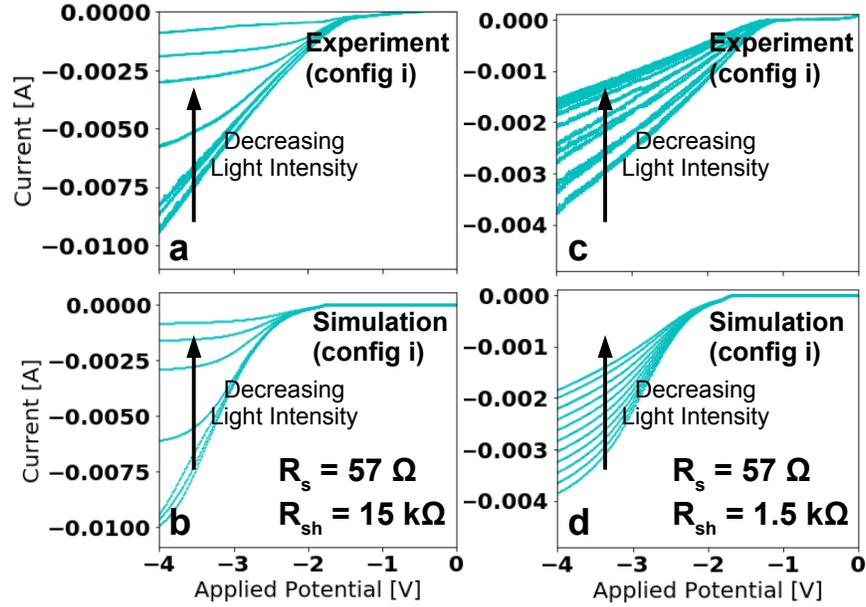

**Figure 4**

*Left:* Experimental (**a**) and simulated (**b**) curves of a photoelectrolyzer with a photoelectrochemical anode featuring high shunt impedance ($R_{sh}$ = 15 kΩ). Simulated curves are generated using equations 5a and 5b and are shown for illumination levels corresponding to 6, 13, 25, 55, 95, 99 and 100% $I_{sc}$; $I_{sc}$ = 11 mA. *Right:* Experimental (**c**) and simulated (**d**) curves of a photoelectrolyzer featuring a low shunt resistance (simulated $R_{sh}$ = 1500 Ω) due to a thicker Ni catalyst layer. Reduced shunt impedance is shown to be a primary cause of increased dark currents in devices where Ni catalyst thickness significantly exceeds ~ 2 nm. Simulations shown for illumination levels corresponding to 0 – 100 % $I_{sc}$; $I_{sc}$ = 4.4 mA.



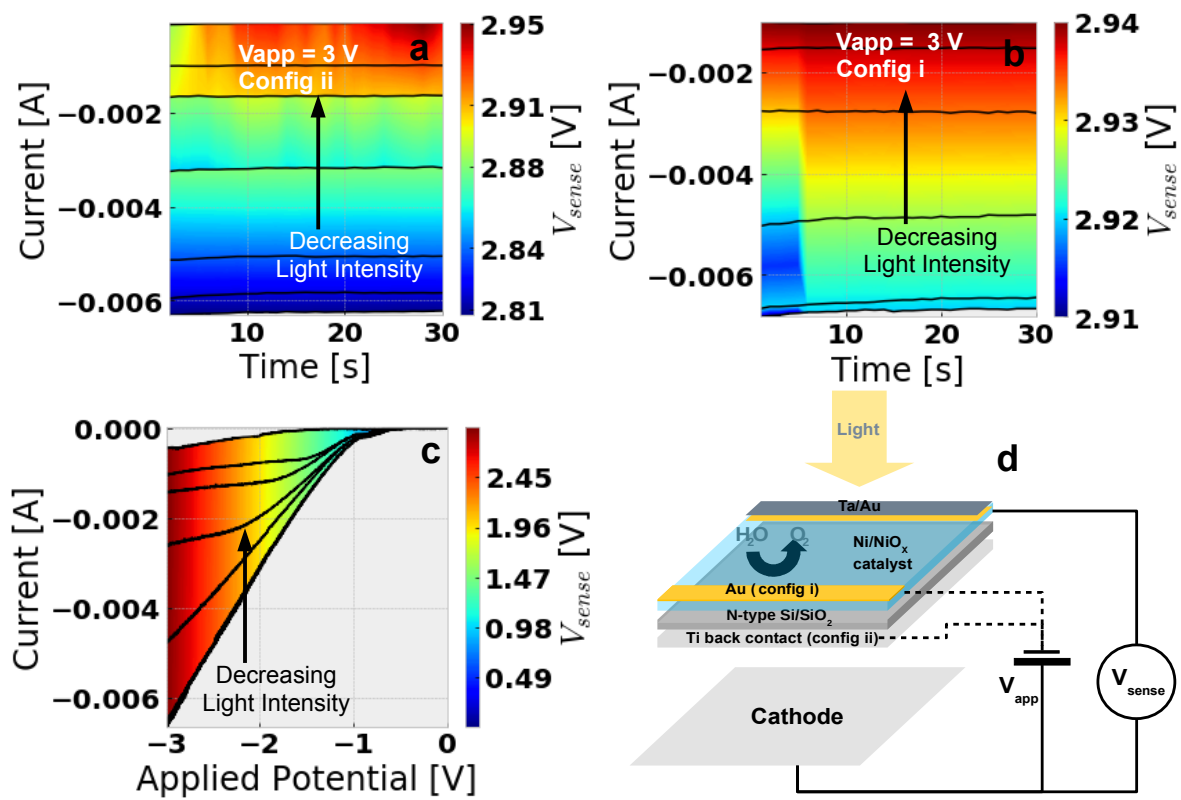

**Figure 5**
(**a**) Changing light intensity for an experimental cell in configuration (ii). Potential is applied (3 V) between the anode back-contact and cathode. A secondary potentiostat channel ($V_{sense}$) monitors the potential between the cathode and anode front-contact. Here, changing light intensity causes significant potential variations (up to 140 mV over a range of 6 mA) between cathode and the anode semiconductor-liquid junction. (**b**) Changing light intensity for an experimental cell in configuration (i). A potential is applied between the photoanode front-contact and cathode (3 V). Fluctuations in $V_{sense}$ are found to be significantly less when biasing between the cathode and anode front-contact and varying current with changes in light intensity (< 30 mV over a range of 6 mA). (**c**) Contours generated through multicontact sensing with voltammetry with the cell in configuration (i) demonstrate the possibility of stabilizing cell potential as current is varied with light. (**d**) Cell electrical connectivity for $V_{app}$ and $V_{sense}$.



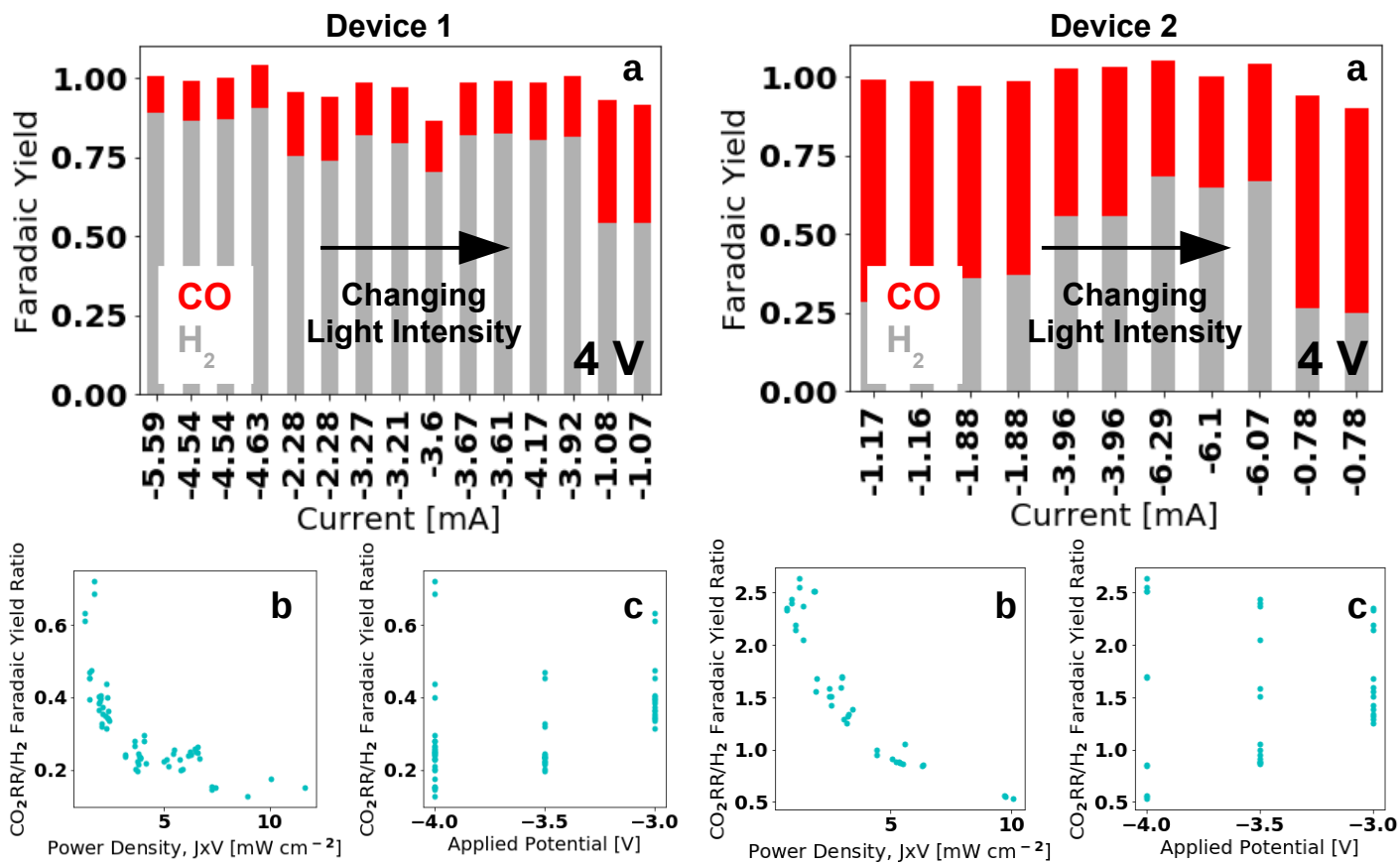

**Figure 6**

(a) product detection for $CO/H_2$ evolution where $V_{app} = 4$ V. Changes in current are facilitated through changes in light intensity incident on the Ni/Si photoanode. Monotonic decreases in the $CO/H_2$ faradaic yield ratio are observed with decreases in light intensity at this single value of the applied cell potential. (b) Changes in power density as a function of $CO/H_2$ ratio, collected over multiple current densities, at 3, 3.5 and 4 V. (c) Changes in the product distribution are shown to be not uniquely dependent on cell potential, as multiple faradaic yield ratios may be observed at a single value of the applied potential, depending on the current density.



# An Expansion of Polarization Control Using Semiconductor-Liquid Junctions


Peter Agbo[1,2,3]*

Chemical Sciences Division[1], Liquid Sunlight Alliance[2], Molecular Biophysics & Integrated Bioimaging Division[3], Lawrence Berkeley National Laboratory, Berkeley CA 94720

pagbo@lbl.gov


## Supporting Information

**Index**





## S1. Experimental & Analytical Methods
*Photoanode Fabrication*
Silicon <100> wafers, with reported resistivities of 0.004-0.007 Ω-cm (used for high conductivity, light-independent controls) and 0.1-0.5 Ω-cm (used for light-responsive treatments) were purchased from University Wafer and used as received (with native $SiO_2$). Before use, wafers were cleaned of possible surface contaminants and adventitious carbon by successive sonication in ethanol, acetone, and finally Millipore water. After cleaning, 40 nm Ti was then deposited on the back (unpolished) side of wafers using an Angstrom Engineering Nexdep e-beam metal evaporator. Deposition was performed at a rate of 1.8 Å $s^{-1}$. Ni layers were deposited on Si (polished side) by loading wafers into the AJA sputter depsition system and pumping down to a pressure of $10^{-6}$ Torr. Ni deposition then proceeded under 3 mT argon at a 35 mm deposition distance from the target. Target power was 150 W with a typical target voltage of 175 V. Sample rotation was used to ensure isotropic deposition of Ni on the Si substrate. Deposition proceeded for 26.7 seconds to yield an approximate Ni layer thickness of 2 nm (an 0.75 Å $s^{-1}$ Ni deposition rate was measured by in-situ quartz crystal microbalance sensing). Modified wafers were scribed into ~4 $cm^2$ chips and then mounted for deposition of the Au busbar front contact. In preparation for this, cut wafers were mounted and masked such that the surface and edges of the wafers were covered (edges were protected using kapton tape to avoid shorting between front and back contacts), leaving only a thin (~2-3 mm wide) band along one side of each wafer exposed. These wafers were then loaded into the AJA sputter deposition and a 100 nm Au band was deposited on top of the Ni/Si surface along the unmasked region. A secondary busbar comprised of 100 nm Ta/100 nm Au was also deposited on select chips. These Ta/Au busbars were used as wet contacts for potential sensing (no epoxy insulation). Afterwards, tantalum foil strips were cut and established as ohmic contacts to the Ti layer and Au busbar using silver epoxy. Ta contacts were tested with a multimeter after each stage of silver epoxy cure to ensure low contact resistances. After curing, EPO-TEK 302-3M two-part epoxy (Epoxy Technology) was used to insulate the front and back contacts as well as the cut edges of the Si wafers. Following the curing of 302-3M, a final epoxy layer consisting of a higher viscosity mixture of ~3:1 302-3M : Hysol 9460 (Loctite) was carefully applied around the edges of the Si chips as a final layer of protection (Figure S1).

*Cathode Fabrication*
Au/carbon cathodes were fabricated through sputter deposition on Toray carbon paper (Fuel Cell Store) using the AJA radio frequency metal sputtering system. An Au target (99.99% purity, Lawrence Berkeley National Laboratory) was used as the target. The sample plate was rotated to ensure even deposition thickness across the substrates. Chamber pressure was adjusted to 3 mTorr argon and the Au target adjusted to 150 W, with a typical target voltage of 160 V. Samples were exposed to the plasma for in order to establish a 100 nm of Au on the Toray carbon papers (typically ~3 Å $s^{-1}$ deposition rate).

*Cell Fabrication and Assembly*
50 mm x 50 mm cell endplates were fabricated from polymethylmethacrylate (PMMA, McMaster-Carr Supply Company, Santa Fe Springs, CA) in house at Berkeley Lab. The flow channels for supplying the anode and cathode feeds were composed of polyetheretherketone (PEEK) tubes, attached with a two-component epoxy (EPO-TEK® 302-3M) to each endplate.

> 1) Cathode endplate with flow ports
> 2) 50 mm x 50 mm orange silicone gasket (1 mm thick) with 2 cm x 1.1 cm substrate channel
> 3) Ta foil current collector with 2 cm x 1.1 cm substrate channel
> 4) Au/Carbon cathode 2 cm x 1.25 cm (Au layer face up)
> 5) Sustainion membrane
> 6) 50 mm x 50 mm orange silicone gasket (1 mm thick) with 2 cm x 1.1 cm substrate channel.
> 7) Ni/Si/Ti photoanode chip, 2 cm x 2 cm (Ni side up)
> 8) 50 mm x 50 mm orange silicone gasket (1 mm thick) with 3 cm x 3 cm channel.
> 9) 50 mm x 50 mm silicone gasket (3 mm thick), with 4 cm x 4 cm channel
> 10) Anode endplate with flow ports

Following assembly, cells were tightened by torquing Torx screws to 30 inch-oz (Figure S2).



The Ta current collector was produced through laser cutting a 100 μm thick tantalum foil, creating a 0.9 mm wide channel with a length of 17.3 mm. Prior to testing, membranes were stored in 1 M KOH solution, and rinsed in Millipore water immediately before assembly.

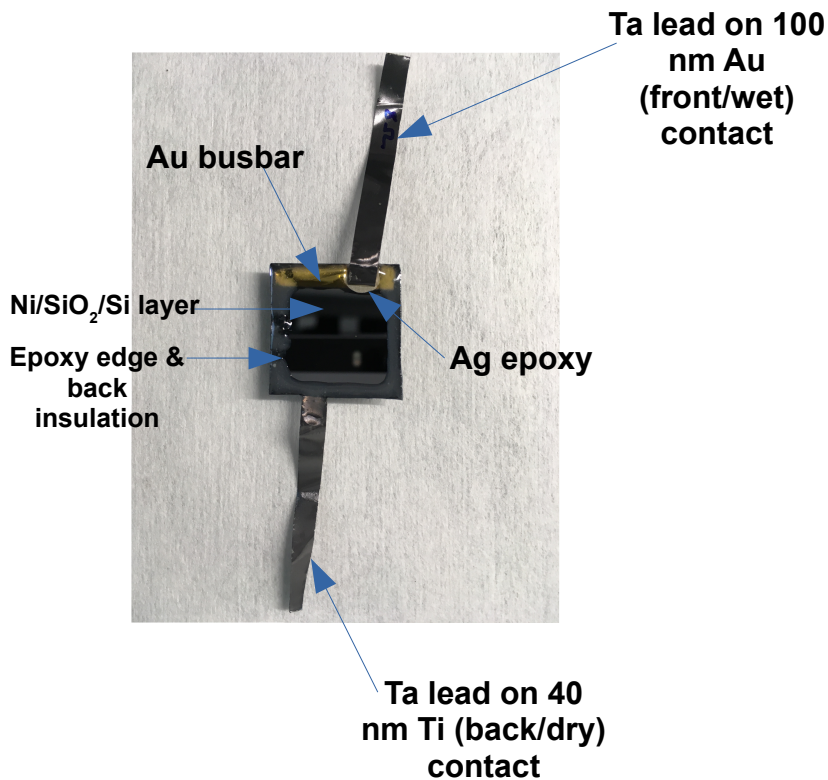

**Figure S1**
Photoanode (counter electrode) fabrication.

*Electrochemistry*
Electrochemical measurements were collected with a Gamry Reference 600+ potentiostat. Assembled, two-electrode cells, acting on an anolyte consisting of 0.65 M lithium borate + 0.35 M potassium borate at pH 9.5, was pumped through the anode volume at a rate of 2.5 ml min$^{-1}$ via peristalsis. Anodes were conditioned by cycling at least 25 times between 0 and 3V full cell potential, with anode illumination such that total cell current at 3 V was between 3 and 5 mA. If necessary, additional cycling proceeded until voltammograms converged. Cell cathodes were flushed with $CO_2$ (Airgas) at a flow rate of 5 sccm. Gas flow rates were controlled using flow meters (0.5-10 sccm resolution) purchased from Alicat Scientific. All applied potentials are referenced vs a full-cell potential, with the anode termination serving as a pseudoreference and the Au cathode as the working electrode. A 150 W xenon lamp (Newport) was used to illuminate the cell photoanode. A peak lamp power of 695 mW at 500 nm was measured using a power meter (Newport) after subtracting influences from ambient light and zeroing the instrument baseline, corresponding to a 27 mW cm$^{-2}$ illumination intensity at the photoanode surface (25.5 cm$^2$ spot size).

Adjustments to incident lamp intensity were made through use of a continously-variable, rotary neutral density filter (Thorlabs), mounted on a motor rack and automated using an arduino microcontroller. Arudino scripts were coded such that the controller turned the filter in arcs of approximately 1/6 radians, to allow for light collection at 6 discrete light illumination levels. This level of radial spacing on the filter was found to provide meaningful current variation for testing current-dependent $CO_2$ reduction, and when used in conjunction with built-in script functions for potentiostat biasing made available in the Gamry software, allowed for full automation of device testing.



*Electrochemical Impedance Spectroscopy*
Electrochemical impedance spectra were collected by application of a -3.0 V applied cell voltage (dark) with a 10 mV perturbation amplitude to a cell. Frequency ranges used for impedance spanned 0.2 Hz – 1 MHz. Data were fit using a Kramers-Kronig fit (8 points per decade), and used to determine the value for the ohmic resistance, $R_\Omega$, which was used for IR correction.

*Gas Chromatography (GC)*
Measurement of gas-phase, $CO_2$ reduction products were determined using an SRI 8610 Gas Chromatograph (MG-5 device configuration). TCD and FID detectors were calibrated using standardized tanks for hydrogen and carbon monoxide at various concentrations acquired from Airgas. Argon (Airgas) was used as a carrier gas, supplied at a 30 ml min$^{-1}$ flow rate, for the GC mobile phase. Data were collected using a modified run sequence, programmed to end following hydrogen and CO peak elution, and restart after a two minute purge sequence (5.5 minute total run time per injection cycle).

*Chemicals & Solutions*
Cell anolytes consisted of potassium hydroxide and lithium hydroxide, purchased from Sigma-Aldrich, were used to pH-adjust solutions of 1 M boric acid to pH = 9.5, generating stock solutions of 1M lithium borate and 1M potassium borate. From these stocks, a composite buffer consisting of 0.35 M $LiBO_4$ and 0.65 M $KBO_4$, was generated and used as the anolyte for all electrochemical cell experiments. All solutions were made in filtered, Millipore (18 MΩ-cm) water.

*Data Analysis*
Faradaic yields of potentiostatic data paired with GC analysis were determined using custom scripts written in the python programming environment.

*Analytical Methods*
Derivation of the analytical model describing polarization behavior of a photoelectrolyzer (equation 5a) was done as described previously, using the following mechanism as a basis[1]:

$$A + S \underset{k_1,\, k'_1}{\longleftrightarrow} [A\text{-}S_{(ADS)}] \underset{k_{ET},\, k'_{ET}}{\longleftrightarrow} [A\text{-}P_{(ads)}] \xrightarrow{k_2} A + P$$

Here, A denotes an electrode active site (in the specific case of anode-limited currents, these are taken as the Ni active site concentrations), S is the substrate concentration (OH$^-$ for base-mediated OER; $10^{-5}$ M); A-S$_{(ads)}$ is the concentration of active site-OH adsorbates, A-P$_{(ads)}$ is the concentration of active site-oxygen adsorbates, and P is the concentration of free $O_2$ product. Evaluation of the resulting system of differential equation yielded by this process in the steady-state limit yields equation 5a, expressing the sum of concentrations of A, A-S$_{(ads)}$ and A-P$_{(ads)}$ intermediates as the total Ni active site concentration $[A]_T$[1]. In the parameters used for modeling, $[A]_T = 10^{14}$ cm$^{-2}$, $k_1 = 6 \times 10^{-5}$ cm$^2$ s$^{-1}$ for the OH$^-$ diffusion constant, and k'1 is assigned a value of $k_1[OH]$. Descriptions for $k_{ET}$ and $k'_{ET}$ are given by the Butler-Volmer equation:

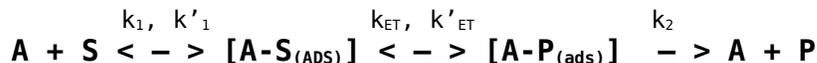

$$k_{ET}(\eta) = k_0 \left[\exp\left(\frac{F\eta\alpha}{RT}\right)\right] \;;\; k_{ET}'(\eta) = k_0\left[\exp\left(-\frac{F\eta(1-\alpha)}{RT}\right)\right]$$

For the purposes of modeling in this study, α = 0.5, F = 96485 C mol$^{-1}$, T = 293 K, and R = 8.3145 J mol K$^{-1}$. Term η denotes the cell overpotential, a variable parameter. The exchange rate, $k_0$, is calculated from an assumed exchange current density typical for nickel-mediated OER under conditions relevant to this work, 10 μA. The rate $k_2$ is pegged to a typical turnover rate typical of heterogeneous catalysts, 0.1 s$^{-1}$. However, sensitivity tests of this parameter showed little influence over polarization behavior in the light-coupled case over several orders of magnitude ($10^{-3} < k_2 < 10^3$ s$^{-1}$ examined), with polarization over the overpotentials tested being primarily controlled by interfacial electron transfer and the limiting photocurrent, $J(\varphi)_{sc}$.



## S2. Photoelectrolyzer Cell Assembly Diagram

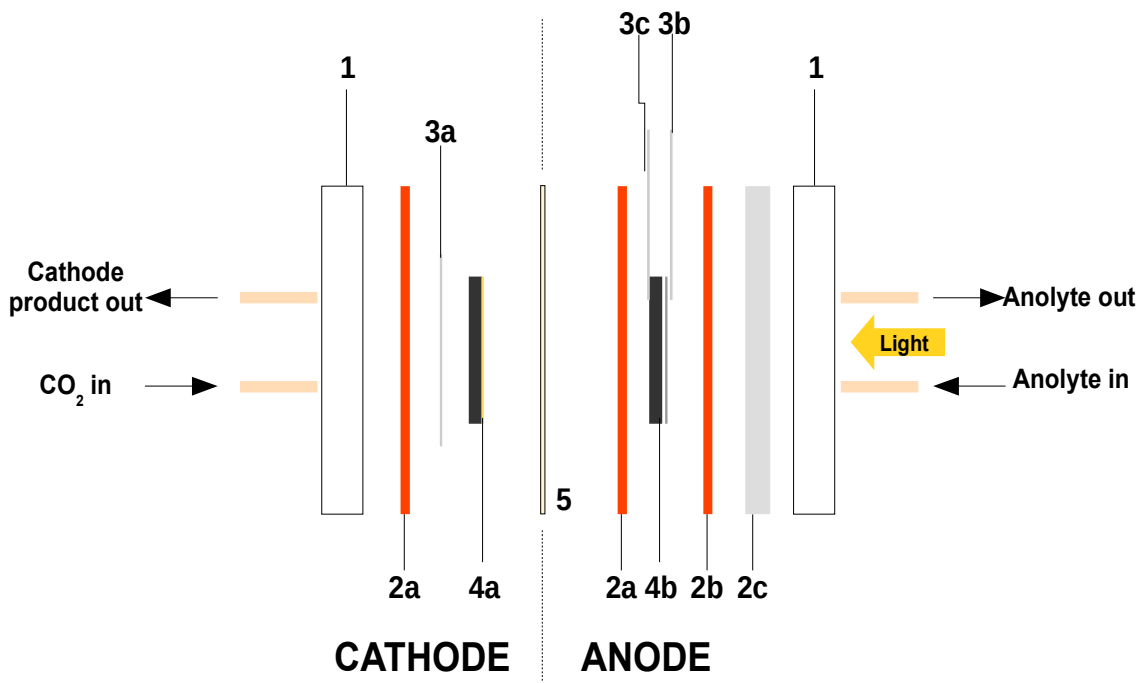

**Figure S2a – Blow-up of electrochemical cell component**
1. 50 x50 mm polyacrylamide endplate
2a. 1 mm silicone gasket, 1 x 2 cm channel
2b. 1 mm silicone gasket, 2 x 2 cm channel
2c. 3 mm silicone gasket, 3 x 3 cm channel
3a. Ta Cathode current collector, 1 x 2 cm channel
3b. Photoanode Ta foil front contact
3c. Photonode Ta foil back contact
4a. Au/Toray carbon cathode
4b. Ti/Si<100>/Ni/NiOx photoanode
5. Ion exchange membrane

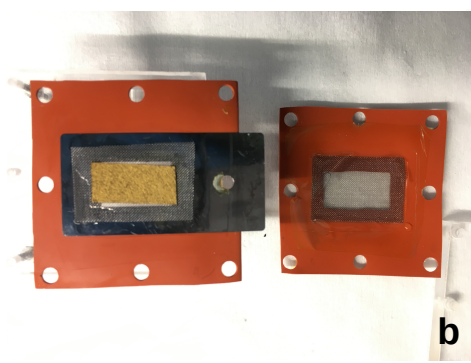
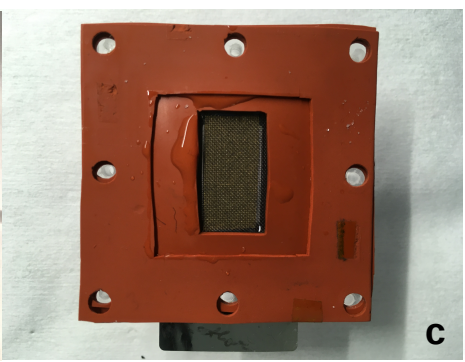
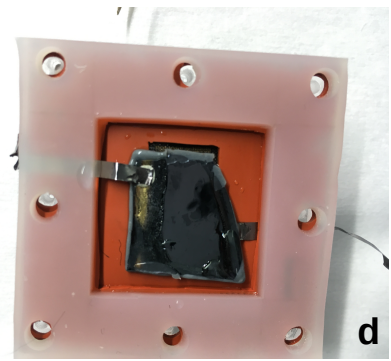

**Figure S2b - Cathode assembly, into page (left):** Au/Toray carbon cathode, Ta mesh current collector (for compression), Ta foil current collector, 1 mm orange gasket, polyacrylamide cathode backplate.

**Figure S2c - Partial anode assembly, into page:** 1 mm orange gasket (2x2 cm Channel); 1 mm orange gasket (1x2 cm channel); Ta mesh compressor; Sustainion membrane; Cathode assembly.

**Figure S2d - Full anode assembly (minus anode backplate), into page:** 3 mm silicone gasket, 1 mm orange gasket (2x2 cm Channel); Photoanode chip with leads between gaskets for sealing; 1 mm orange gasket (1x2 cm channel); Ta mesh compressor; Sustainion membrane; Cathode assembly.



## S3. Polarization Model Sensitivity Analysis

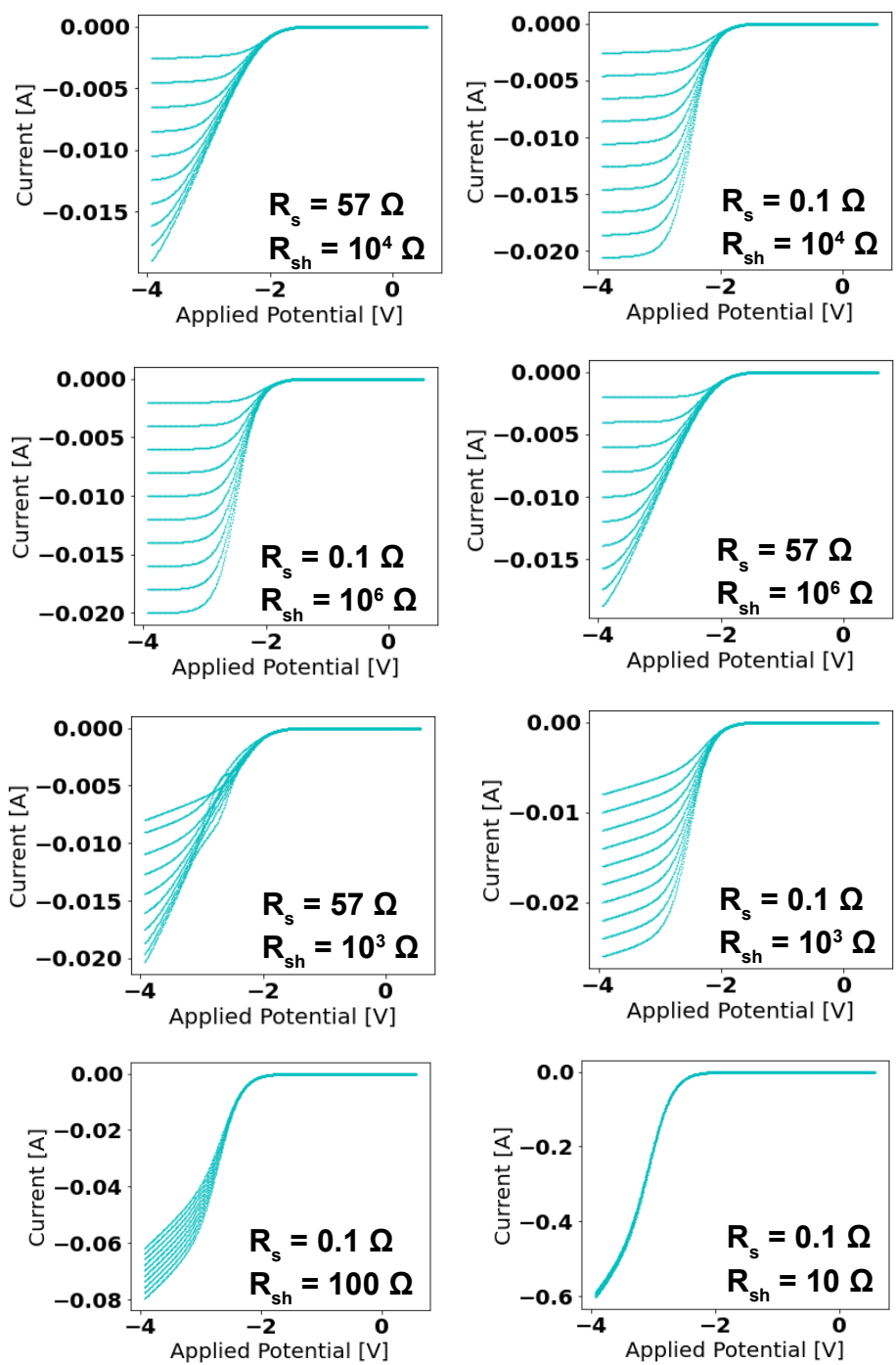

**Figure S3.**
Simulated dependencies of the analytical polarization model (eqs. 4a & 4b) describing photoelectrolyzer device function. Illumination levels from 0 to 100% $J_{sc}$ (= 20 mA) shown for all cases. IR loss effects are incorporated numerically, by recursive calculation of equation 5a over differential potential steps, dV, from 0 to V, with current densities $J(\phi, V_{i-1})$ being used to calculate IR drop at $J(\phi, V_i)$. Step sizes dV are chosen such that $J(\phi, V_i) \sim J(\phi, V_{i-1})$.



## S4. Device Stability

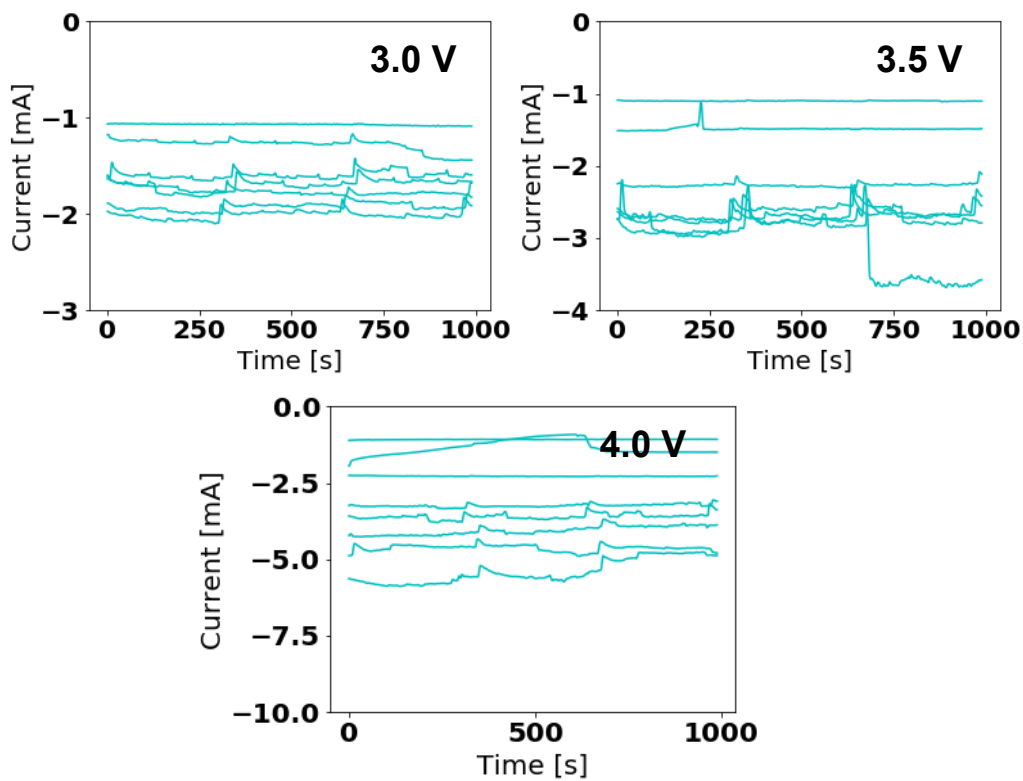

**Figure S4**
Potentiostatic data of device runs over multiple light intensities used for faradaic efficiency data collection (configuration i). Cases of applied cell potentials at 3, 3.5 and 4.0 V shown (cell configuration i). Fluctuations are dominated by periodic bubble cavitation on, and desorption from, the Ni/NiO$_x$ photoanode surface.



**S5. Current-Dependent Product Distributions at 3 and 3.5 V**

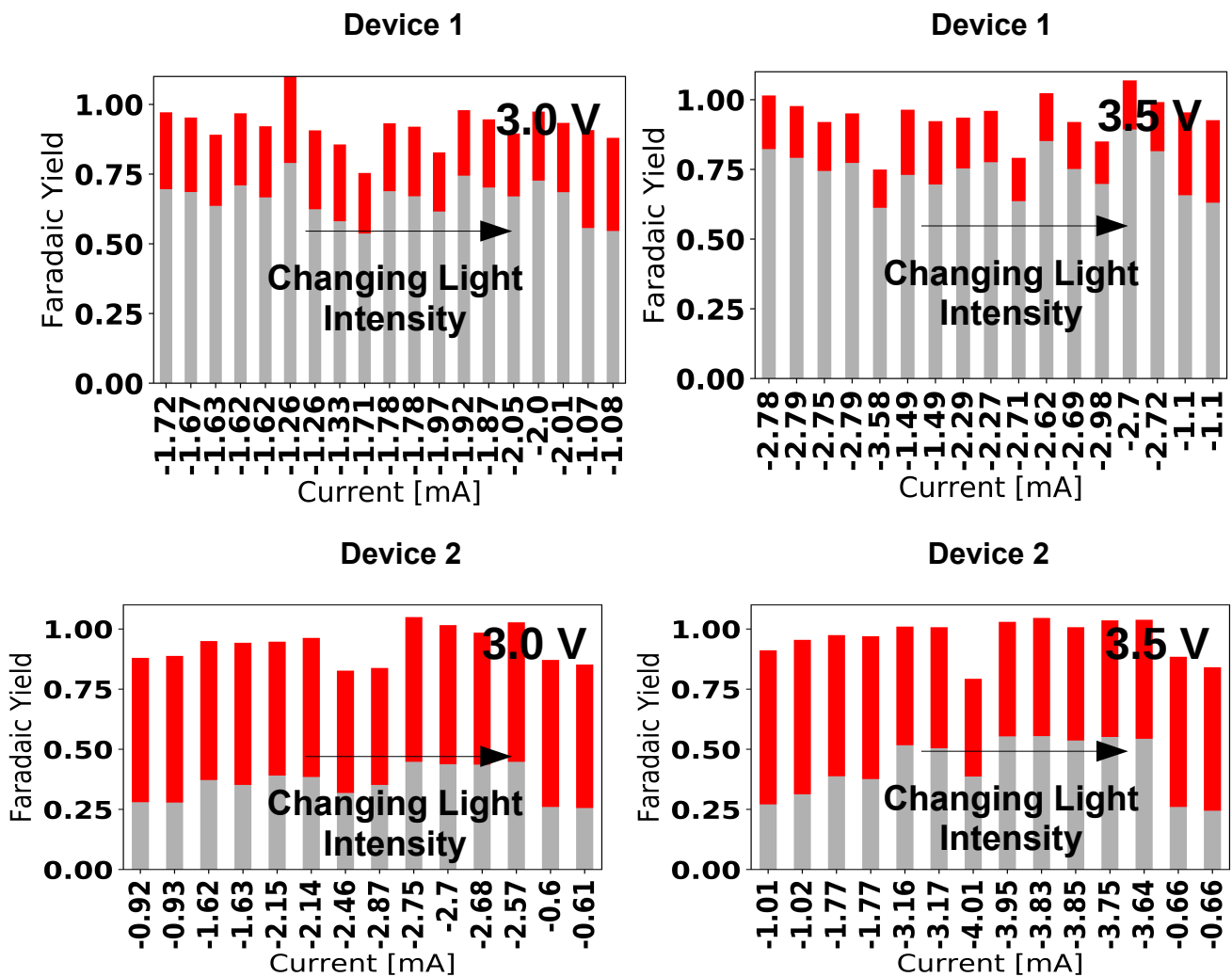

**Figure S5**
Potentiostatic faradaic yields of a representative device runs at 3.0 and 3.5 V (configuration i).



## S6. Product Distribution Dependencies – Power Dependence (Device 1)

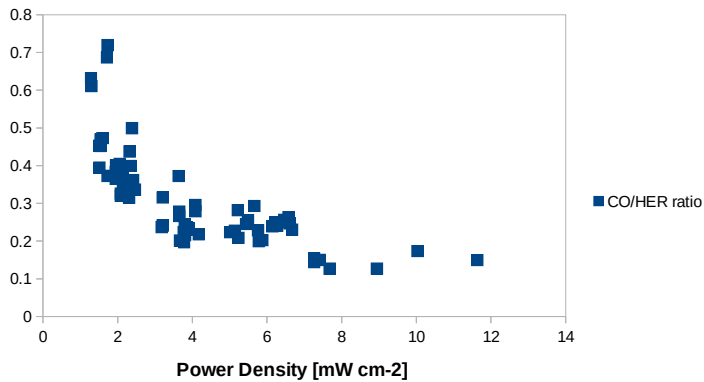
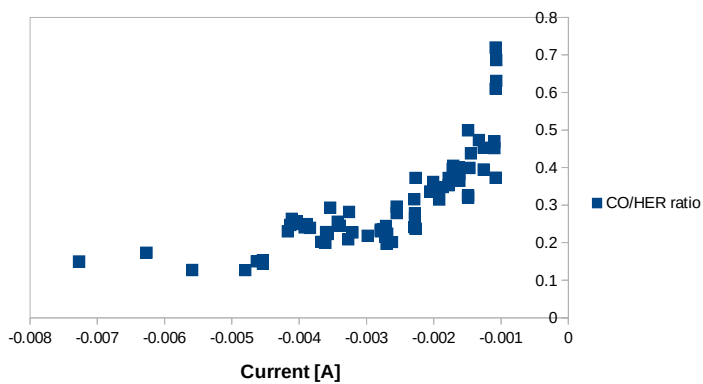
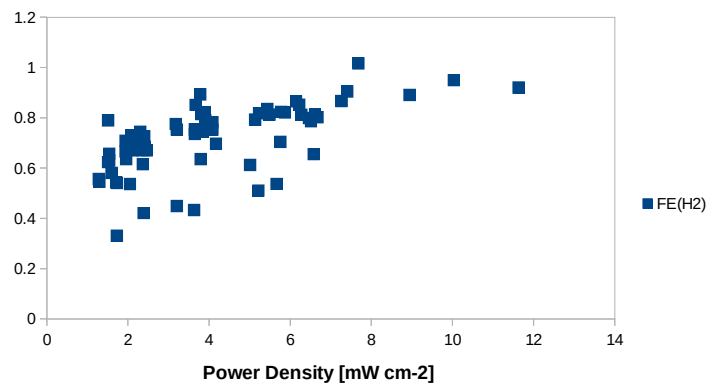
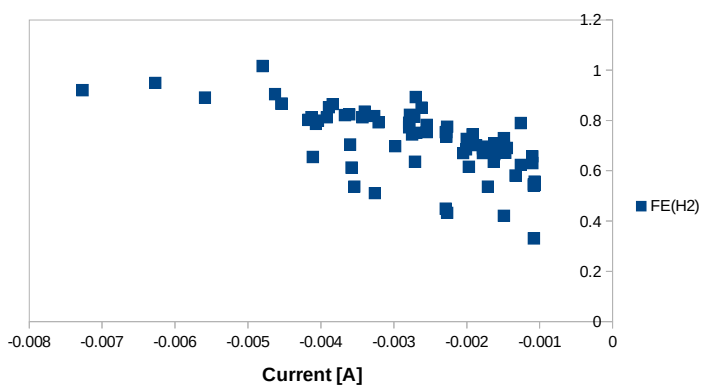
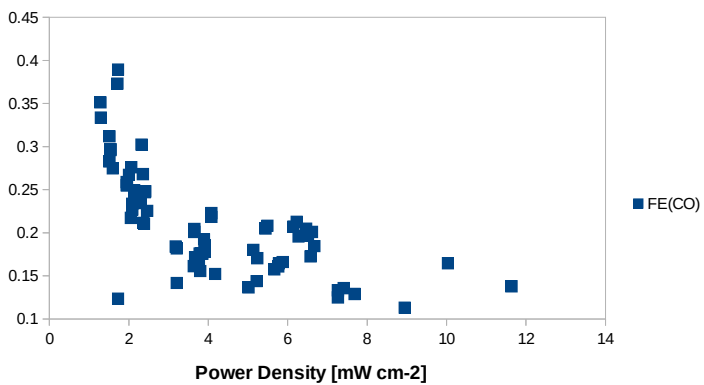
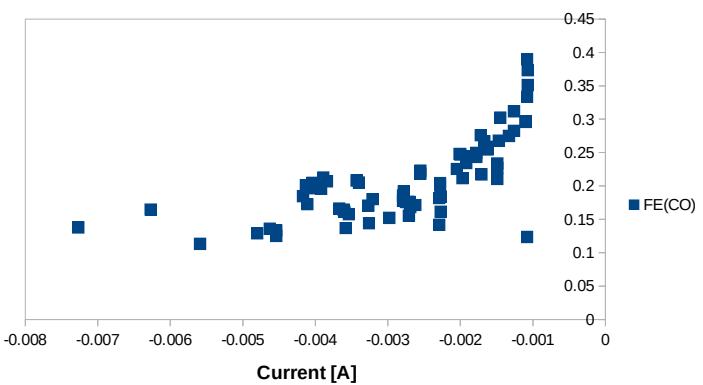



## S7. Product Distribution Dependencies – Potential Dependence (Device 1)

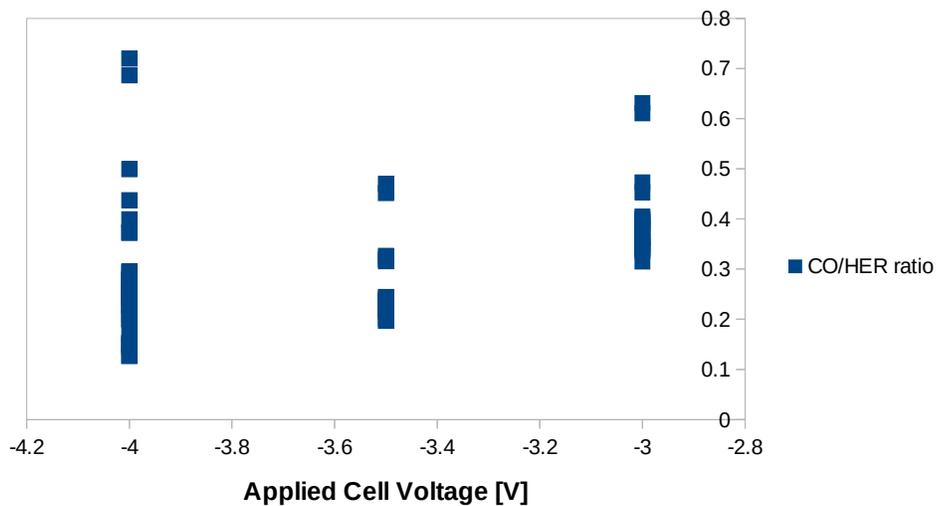

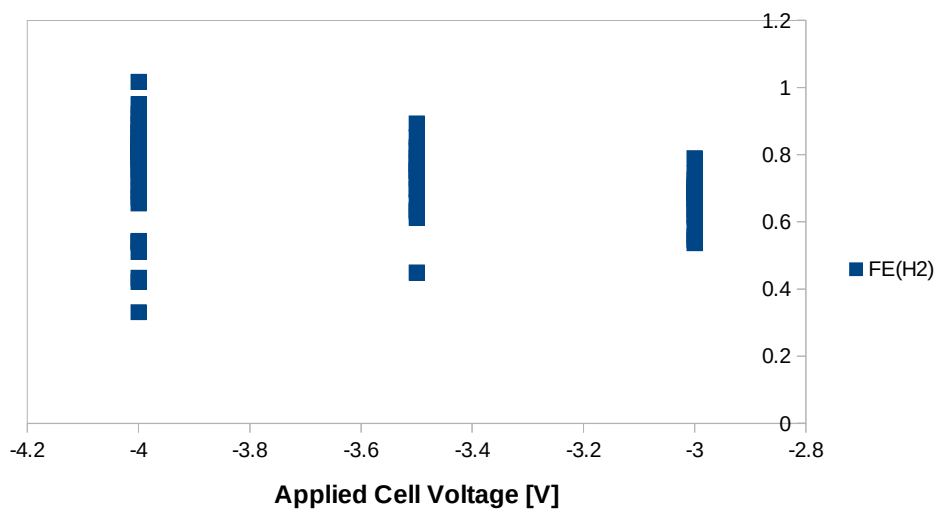

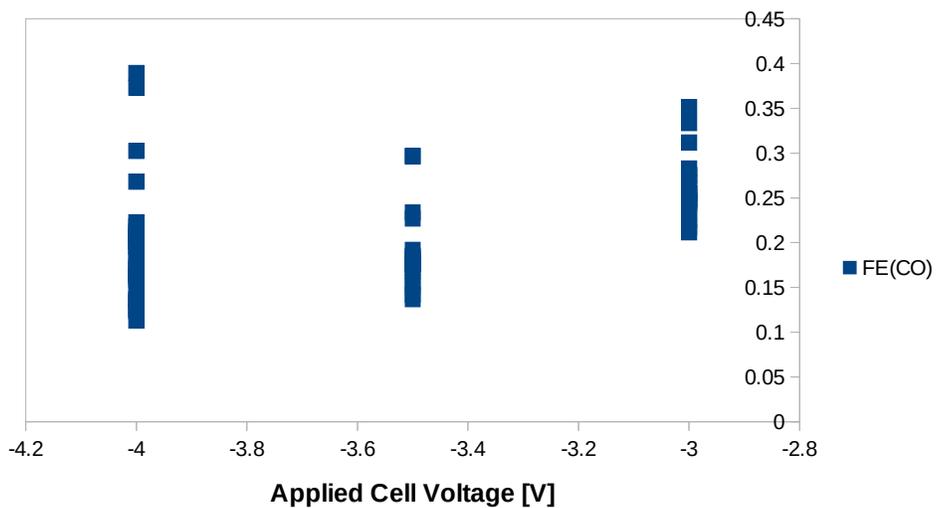



**S8. Product Distribution Dependencies – IR-Corrected Potential Dependence (Device 1)**

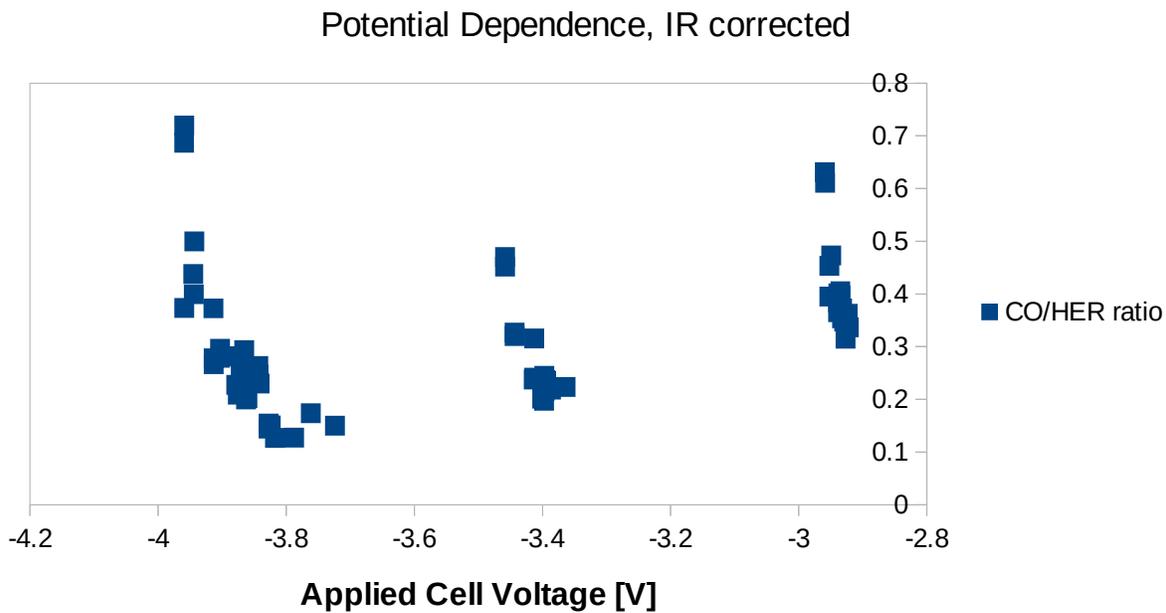

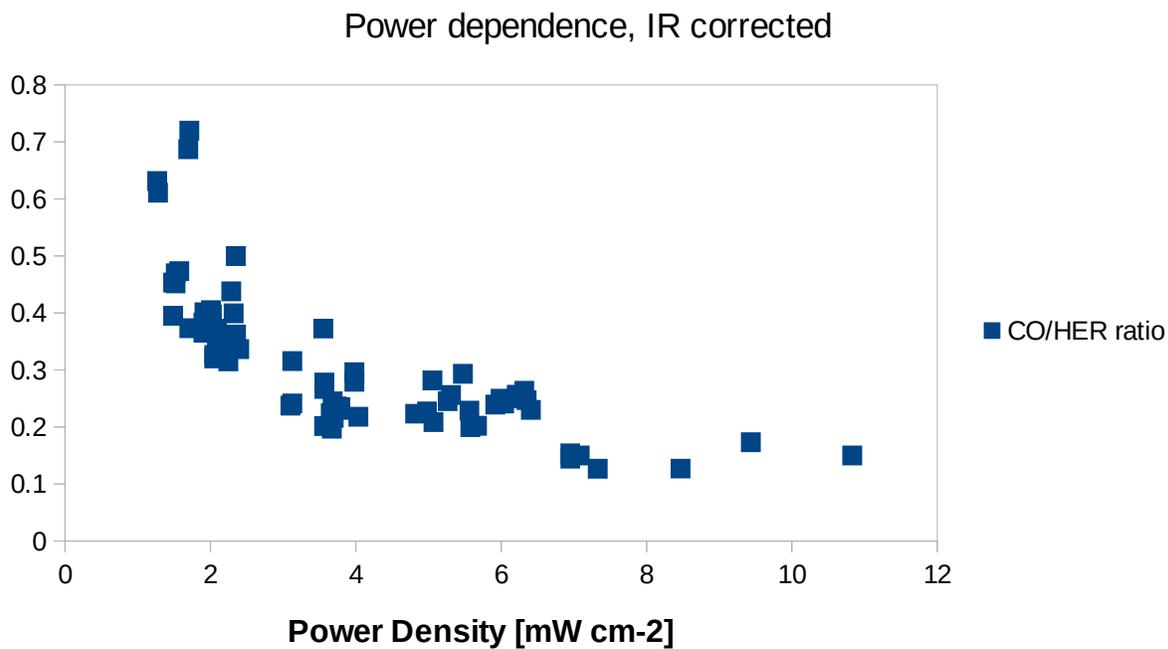

**Figure S8.**
Assembled applied cell potential and power dependencies of CO/$H_2$ ratios after IR correction. Fitting of electrochemical impedance spectra yield an ohmic resistance of 38 ohms (S11).



**S9. Low Shunt Resistance and High Conductivity Si <100> Controls**

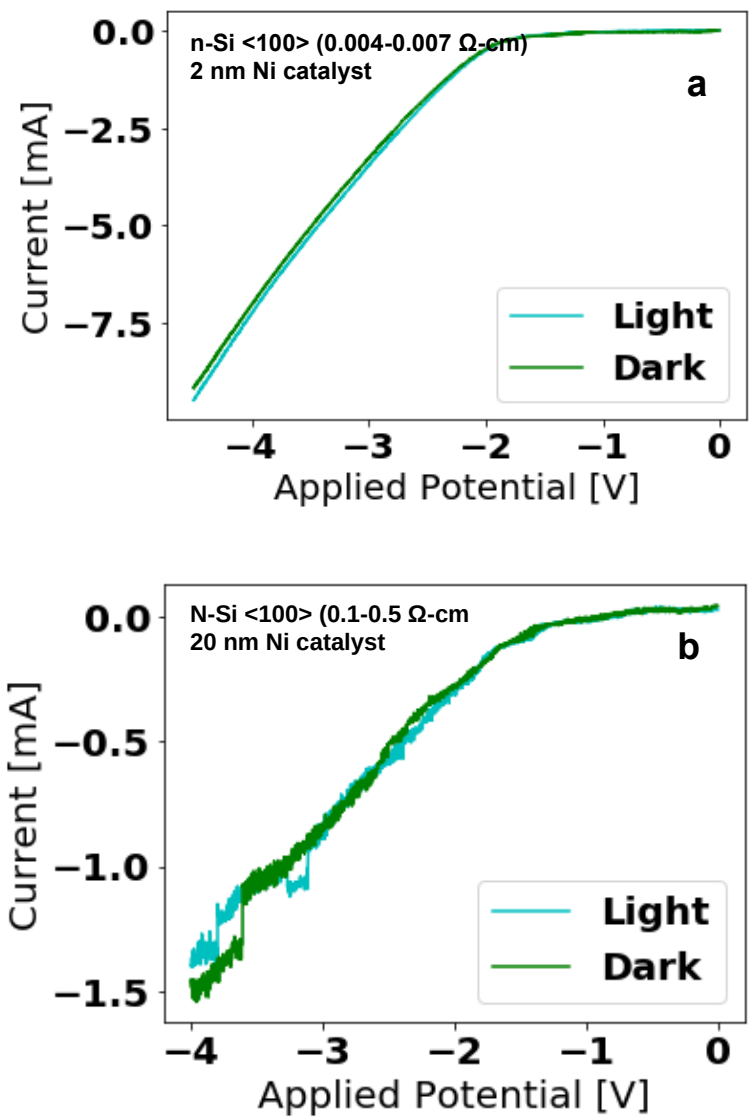

**Figure S9**
a) A representative plot of a cell assembled with front-contacted photoanodes (configuration i) fabricated from high-conductivity Si <100> (0.004-0.007 Ω-cm). High carrier density of n$^+$$^+$-doped Si forces conduction band occupancy by electrons at room temperature, resulting in a material where carrier flow is not responsive to light illumination levels. b) A representative plot of a decoupler cell assembled with a front-contacted photoanode fabricated with a 20 nm Ni layer on the moderate conductivity Si <100> (0.1-0.5 Ω-cm) used for the primary decoupling experiments. Here, thicker Ni catalyst deposition on Si results in a highly conductive metal layer, resulting in a buried semiconductor-liquid junction. As a result, electrolyzer polarization behaves similarly for dark and light conditions, consistent with the polarization model case featuring low values of the shunt resistance (S4).

## S10. Electrochemical Impedance Spectroscopy – Variable Light

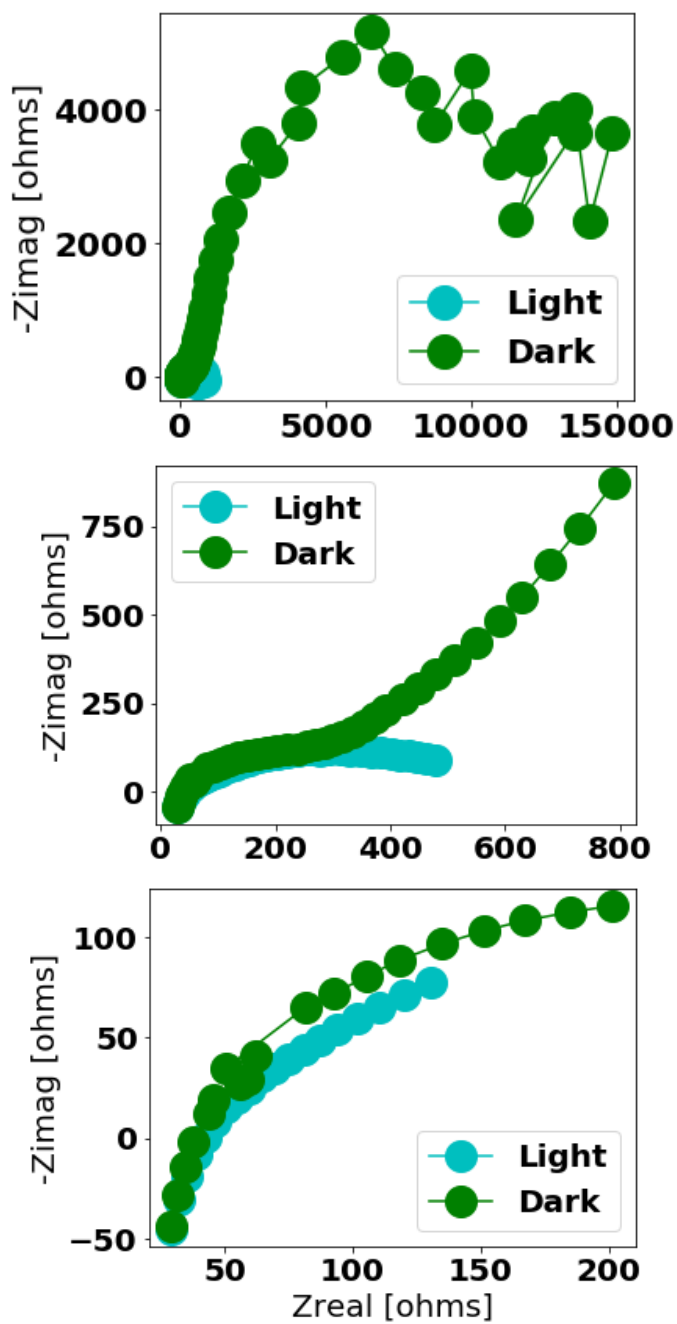

**Figure S10**
Electrochemical impedance spectra, shown at different scales, of a representative 2-electrode EIS measurement of a front-contacted (configuration i) cell. $R_\Omega$ is found to be 37.58 ohms and independent of illumination state.

**S11. Electrochemical Impedance Spectroscopy – Variable R_ext**

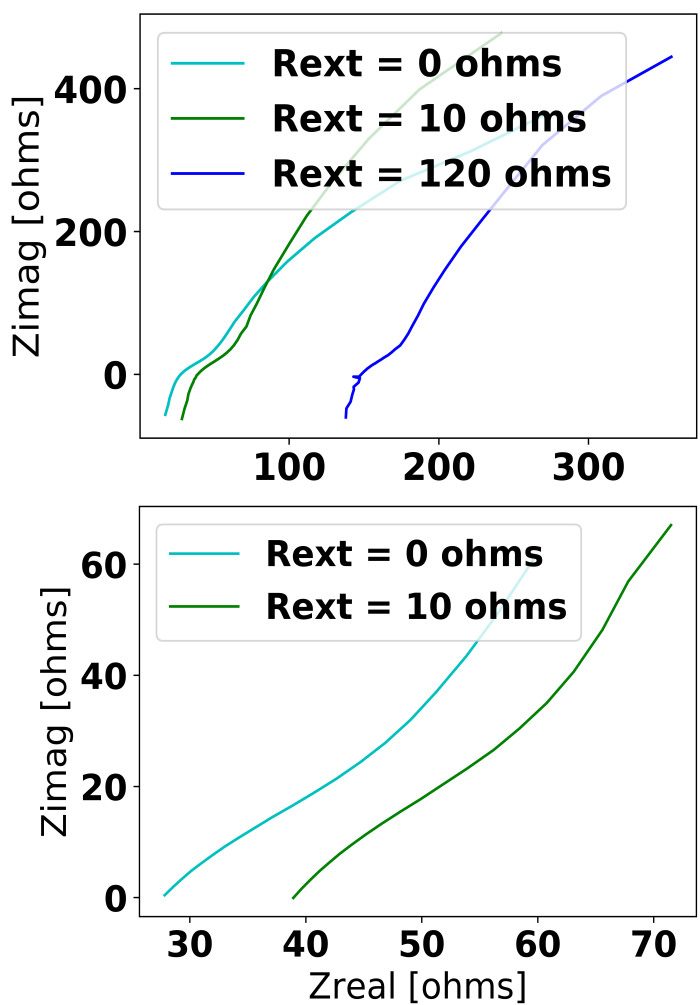

**Figure S11**
Electrochemical impedance spectra collected for a cell with varying values of a resistor ($R_{ext}$ = 0, 10, and 120 Ω) placed in series with the Au contact (between Au contact and potentiostat). Native ohmic resistance of this cell is $R_\Omega$ = 28 Ω (cyan trace). The observed behavior in the high-frequency region of the EIS for these devices are distinct from that of the impedance taken under various light illumination intensities, illustrating that current attenuation is not the mere result of a changing series resistance under variable light intensity.

**S12. Photovoltage Measurement ($V_{Lj}$)**

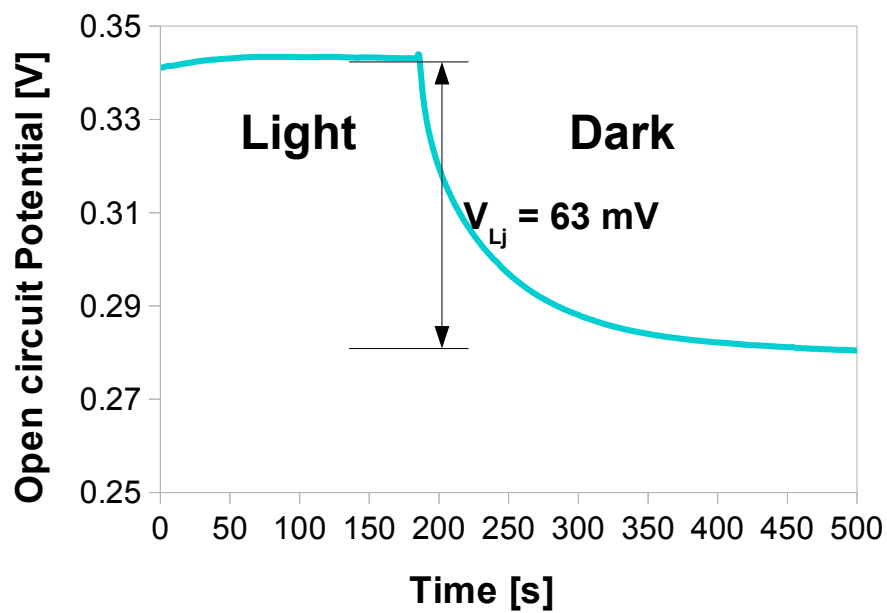

**Figure S12**
Representative photovoltage as measured between the Au contact and cathode in 2-E configuration under dark and light illumination. $V_{Lj}$ = 63 mV.

**References**

(1) Agbo, P. J–V Decoupling: Independent Control over Current and Potential in Electrocatalysis. *J. Phys. Chem. C* **2020**, *124* (52), 28387–28394. https://doi.org/10.1021/acs.jpcc.0c08142.